\documentclass[aps,pra,twocolumn,showpacs,preprintnumbers,amsmath,amssymb,footinbib]{revtex4}
\usepackage{graphicx,epsfig}
\usepackage{placeins}
\usepackage{bm}
\usepackage{dcolumn}
\usepackage{xcolor}
\usepackage{color,soul}
\usepackage[breaklinks=true,colorlinks,citecolor=blue,linkcolor=blue,urlcolor=blue]{hyperref}
\usepackage{xr}

\def\noi{\noindent}
\def\bc{\begin{center}}
\def\ec{\end{center}}
\newcommand{\bea}{\begin{equation}}
\newcommand{\eea}{\end{equation}\noi}
\newcommand{\ber}{\begin{eqnarray}}
\newcommand{\eer}{\end{eqnarray}\noi}
\begin{document}
\title{Parallel refreshed cryogenic charge-locking array with low power dissipation}
\author{X Bian$^{1}$} \email{oums1005@materials.ox.ac.uk}
\author{G A D Briggs$^{1}$}
\author{J A Mol$^{2}$} \email{j.mol@qmul.ac.uk}
\affiliation{$^{1}$Department of Materials, University of Oxford, Oxford, OX1 3PH, UK\\
\\$^{2}$ School of Physics and Astronomy, Queen Mary University, London, E1 4NS, UK}

\begin{abstract}
To build a large scale quantum circuit comprising millions of cryogenic qubits will require an efficient way to supply large numbers of classic control signals \citep{Almudever2017,Franke2019,Vandersypen2017,Hornibrook2015,Reilly2015}. Given the limited number of direct connections allowed from room temperature, multiple level of signal multiplexing becomes essential. The stacking of hardware to accomplish this task is highly dependent on the lowest level implementation of control electronics \citep{Franke2019}, of which an open question is the feasibility of mK integration. Such integration is preferred for signal transmission and wire interconnection, provided it is not limited by the large power dissipation involved. Novel cryogenic electronics that prioritises power efficiency has to be developed to meet the tight thermal budget. In this paper, we  present a power efficient approach to implement charge-locking array. As opposed to conventional approaches, where the power dissipation grows superlinearly with the total number of charge-locking units (quadratic growth with 1-dimensional addressing and to the power of $\frac{3}{2}$ with 2-dimensional cross-bar addressing as will be shown), our charge-locking scheme approaches linear power dissipation at large scale. The reduced power dissipation results from the parallel recharging method employed, which greatly decreases the number of switchings involved. To benchmark the power efficiency, we evaluate the power dissipation required to maintain $2^{14} \times 2^{14} \simeq 2.6 \times 10^{8} $ charge-locking units. As compared with serially refreshed charge-locking array with cross-bar addressing, our parallel refreshed charge-locking array shows more than 5000 times reduction in power dissipation and only dissipates 11 ~$\mu$W per kHz refreshing rate (assuming transistor gate size of 10~nm $\times$ 14~nm). Such a low power dissipation is compatible with the 1 mW cooling power available at 100 mK for large dilution fridges. We envision this highly efficient charge-locking scheme will lead to integrated classical control electronics for cryogenic quantum technologies.
\end{abstract}
\maketitle

\section{Introduction}
\label{Introduction}
Semiconductor spin qubits based on gate defined quantum dot (QD) devices are promising candidates for building a universal quantum computer. They have the potential to reach high-level integration due to their small physical size and their compatibility with established semiconductor production processes. Although tremendous progress has been achieved over the last twenty years\citep{Petta2005,Veldhorst2014,Muhonen2014,Veldhorst2015,Watson2018,Zajac2017,Huang2019,Yang2020,Petit2020, Hendrickx2020, Hendrickx2020a}, a feasible path towards building a large scale quantum circuit based on QD devices remains to be established. The difficulty in scaling up lies not only in interconnecting (coherently coupling) between different QDs \citep{Veldhorst2015,Watson2018,Mi2017,Samkharadze2018}, but also in interfacing with classic control signals (e.g. DC voltage, DC pulse, microwave pulse) that are required to operate (define, control and measure) QDs \citep{Vandersypen2017,Almudever2017,Franke2019,Reilly2015,Hornibrook2015}. Classical control signal interfacing is conventionally supplied from room temperature electronics directly. Such a direct approach, albeit with high flexibility and simplicity, cannot be sustained as it depletes resources (e.g. cooling power, physical space) very quickly. A more efficient way to interface with classic control signals has to be implemented to facilitate upscaling.

A viable interfacing solution for scaling up quantum circuit requires fine tolerance levels for control signals and precise tuning to operate each QD. Large numbers of individually tunable signals have to be simultaneously maintained. Heat load and electrical noise have to be managed. Two different paths have been pursued in parallel towards developing a solution. One aims to minimise the variation of device characteristics, so that shared control lines can be used \citep{Li2018}. High-throughput electrical characterisation capability at cryogenic temperature can accelerate process optimisation \citep{Pillarisetty2019,Wuetz2020,Pauka2019}. Progress has been achieved in fabricating highly uniform double QDs\citep{Borselli2015}, but the required level of uniformity remains to be achieved for a large array of QDs \citep{Zajac2016}.

An alternative approach is to employ signal multiplexing circuitry and an array of charge-locking units \citep{Veldhorst2017,Vandersypen2017}. Each charge-locking unit is made up of a capacitor and a switch. The capacitor can be charged up to a static voltage when the switch is set to ON state and can hold that voltage for a period of time when the switch is set to OFF state. To maintain a voltage level, each capacitor has to be periodically recharged. A few proof-of-concept demonstrations \citep{Pauka2019a,Xu2020,Puddy2015,Schaal2018,Schaal2019} have been reported so far. The schemes reported in \citep{Pauka2019a} and \citep{Puddy2015} can be extended to charge-lock a large number of DC signals. The architecture suggested in \citep{Schaal2019} is also scalable by frequency multiplexing. In \citep{Pauka2019a}, signal multiplexing is accomplished by a cryogenic digital finite state machine (FSM) and digital clock to operate, which together contribute significant power dissipation, leaving sufficient cooling power for only  $\sim$1000 charge-locked signals to be maintained. Further scaling  will require some functionalities to be implemented at higher temperature to reduce the cryogenic heat load. An analog multiplexer (MUX) can control an array of charge-locking units using a multiple level selective gating (MLSG) method \citep{Puddy2015}. The number of multiplexed outputs depends exponentially  on the number of control lines, and can rely on small number of control lines (reducing direct heat flow) supplied from control electronics implemented at higher temperature (mitigating heat load) to operate. The dependence of power dissipation on the total number of charge-locking units remains to be analysed.

In this paper, we will first analyse the power dissipation involved in operating this multiplexed 1-dimensional (1D) charge-locking circuitry. We will show that the power dissipated to perform periodic recharging is dominated by the MUX and grows quadratically with the total number of charge-locking units. Then the power dissipation analysis is extended to 2-dimensional (2D) charge-locking array implemented with conventional cross-bar addressing approach (see for example \citep{Veldhorst2017,Vandersypen2017}), which also shows superlinear dependence (to the power of $\frac{3}{2}$) on the total number of charge-locking units. Certainly, the accelerated growth rate of power dissipation with the size of charge-locking circuitry is not desired. Faced with this issue, we will propose a different scheme to charge-lock DC voltages at scale, where the signal lines for charging capacitors are multiplexed instead. A key improvement in our approach is that the power dissipation grows (very close to) linearly with the total number of charge-locking units (for both 1D and 2D addressing). The linear dependence results from the capability of our approach to perform periodic recharging in a parallel manner. Apart from the power dissipation aspect, improvement can also been achieved in the total time required to perform periodic recharging. This total recharging time is another critical aspect that can limit the size of the charge-lock circuitry. Lastly, we will generalise the MLSG method itself. As compared with base-2 multiplexing scheme, base-4 multiplexing scheme is as efficient in terms of scaling up the total number of multiplexed outputs, however will lead to further reduction in power dissipation.

\section{Power dissipation analysis of conventional approach}
\label{PDA}
\begin{figure*}[h]
\includegraphics[width=0.98 \linewidth]{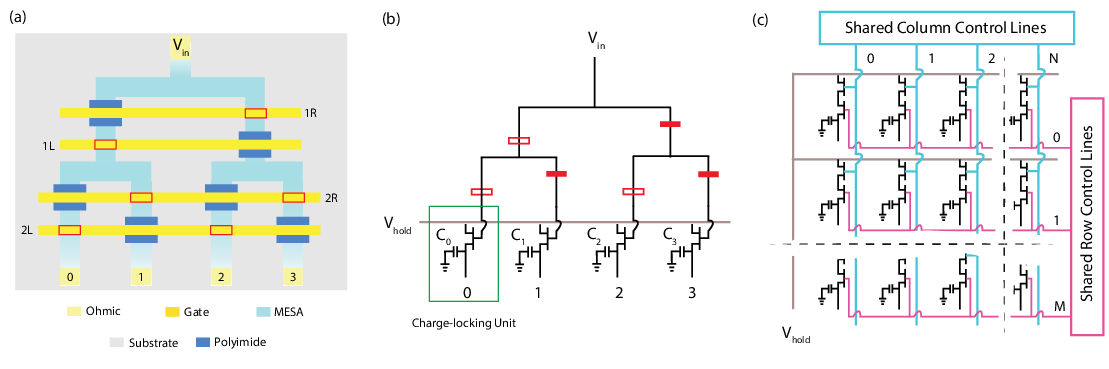}
\caption{(a) Schematics of a 2-level base-2 MUX, which relies on MLSG method to route DC signals. (b) Circuit schematics showing the MUX is used to drive the 4 transistors (i.e. switches) of charge-locking units, which is the approach taken in \citep{Puddy2015} and is the 1-dimensional equivalent of the cross-bar 2-dimensional addressing approach as proposed in \citep{Veldhorst2017}. All the charge-locking units share a same input $V_{hold}$ for charging, such that each charge-locking unit can only be recharged in serially. In the circuit schematics, addressing gates are represented by horizontally aligned rectangles. Hollow red rectangles represent the addressing gates that are not activated and solid red rectangles represent the addressing gates that are activated. Charge-lock unit 0 is selected here. (c) Circuit schematics showing the 2-dimensional charge-locking array with the cross-bar addressing approach. Each charge-locking unit consists of a holding capacitor and a pair of transistors. The pair of transistors are controlled by shared column/row control line respectively, such that $N$ column and $M$ row control lines in total can address $N \times M$ charge-locking units. Each charge-locking unit can only be recharged sequentially. 
\label{CL-Schematics}}
\end{figure*}

Before presenting the detailed analysis on power dissipation, we will briefly explain the basic operation principle of the multiplexed charge-locking approach reported in \citep{Puddy2015}. It is worth explaining here this architecture is essentially the 1-dimensional equivalent of the cross-bar addressing suggested in \citep{Veldhorst2017,Vandersypen2017}. Figure \ref{CL-Schematics}a shows the schematics of a 2-level base-2 analog MUX based on the MLSG method. Physically, base-2 signal multiplexing corresponds to each channel (MESA) at current level being split into two channels at next level. The channel conduction is through the 2-dimensional electron gas (2DEG), which is normally on and can be suppressed by applying negative voltages to addressing gates. The selective channel gating is achieved by using polyimide to modulate gate capacitance. At each level, a pair of addressing gates are complementary, of which one controls the odd channels and the other controls the even channels. Therefore, each output can be individually selected by activating (applying negative voltage) a different combination of addressing gates. For example, output 0 is selected by activating addressing gate 1R/2R and output 1 is selected by activating addressing gate 1R/2L. To implement multiplexed charge-locking, each multiplexed output is connected to the gate of a transistor, which acts as the switch of a charge-locking unit. All the charge-locking units share a same input signal line $V_{hold}$ for charging and the circuit schematics is shown in Figure \ref{CL-Schematics}b. The detailed process to initialise each charge-locking unit to hold a different voltage is described in \citep{Puddy2015} and will not be repeated here. We will focus on the process to perform periodic recharging. In the charge-locked state, all the multiplexed outputs are also charged-locked to hold a static voltage of $V_{g}$, which keeps the transistor of each charge-locking unit in OFF state and requires 2$V_{g}$ to be applied to all the MUX addressing gates. $V_{g}$ is determined by the most negative voltage to be charge-locked. To recharge a specific charge-locking unit (e.g. charge-locking unit 0), the corresponding multiplexed output is connected back to the MUX input (e.g. by deactivating 1L/2L, thus only 1R/2R are activated) and the input $V_{in}$ is set to 0V. The transistor is thus turned on and the charge-locking unit can be recharged from the shared input $V_{hold}$. After recharging, the MUX input $V_{in}$ is set back to $V_{g}$ to turn off the transistor. All the deactivated addressing gates are then set back to 2$V_{g}$ to put all the multiplexed outputs in charge-locked state. Same procedure can be repeated to recharge all other charge-locking units.

The power dissipated in periodically recharging the charge-locking circuitry has three components. First is the power consumed to recharge the capacitors that hold the static voltages (referred as holding capacitor $C_{H}$). Second is the power dissipated in driving the switches of the charge-lock units. The switch is implemented as a transistor, such that the power is essentially dissipated in charging/discharging the transistor gate (referred as transistor gate capacitor $C_{g}$). Third is the power dissipated in the signal multiplexing circuitry. Similar to the second contribution, the power consumed to operate the MLSG based MUX is essentially dissipated in charging/discharging the MUX addressing gates. The first contribution is constrained by the voltage resolution required (hence the minimum $C_{H}$) and the voltage drift $\delta V_{H}$ allowed (i.e. the voltage swing involved in recharging). It has been shown previously in \citep{Xu2020} that the second contribution can be orders of magnitude larger than the first contribution. However, the comparison is made with $~\mu$m size transistor. We will strengthen this point here by showing that even if the transistor is implemented with the most advanced technology, the second contribution still dominates.

As the first two contributions are required for each single charge-locking unit, it is enough to compare them for one charge-locking unit. The first contribution $P_{H}$ can be expressed as
\begin{equation}
P_{H} = C_{H}\delta V_{H}^{2}f_{c},
\end{equation}
where $f_{c}$ is the periodic recharging frequency. The voltage resolution is set by either the charge discreteness $e/C_{H}$ or the thermal noise level $\sqrt{\frac{KT}{C_{H}}}$. To reach a voltage resolution of 1$\mu$V at a temperature of 100 mK, the holding capacitor $C_{H}$ has to be larger than 1.4 pF, which is set by the thermal noise. With the conventional planar technology of 10 fF/$\mu$m\textsuperscript{2}, it might be 14 $\mu$m $\times$ 10 $\mu$m.

Similarly, the second contribution $P_{g}$ can be expressed as
\begin{equation}
  \begin{split}
P_{g} = C_{g}\delta V_{g}^{2}f_{c}\\
  \end{split},
\end{equation}
where $\delta V_{g} = V_{g} - 0 = V_{g}$ is the voltage swing involved in turning on/off the transistor and $C_{g}$ is the transistor gate capacitor. $\delta V_{g}$ is largely determined by the voltage level required to operate QD and is on the order of $\sim$ 1V \citep{Vandersypen2017}. If the voltage drift $\delta V_{H}$ to be compensated is 10 $\mu$V, the transistor gate has to be smaller than 0.14~nm $\times$ 0.1~nm to make $P_{g}$ smaller than $P_{H}$, which certainly is not possible even with the most advanced technology. Suppose the transistor is implemented with a size of 14~nm $\times$ 10~nm, $P_{g}$ is then at least 10\textsuperscript{4} times larger as compared with $P_{H}$. As the transistor gate decreases, the threshold voltage will increase, thus the voltage $V_{g}$ required to turn off the transistor will increase and the power dissipation will be even larger. This will be neglected in the following analysis, where we will approximate the total power consumption as the sum of the second and the third contributions.

For 1D charge-locking circuitry that consists of N charge-locking units being controlled by a K-level base-2 MUX (i.e. $N= 2^{K}$), the energy dissipated to sequentially switch on/off the transistor of each charge-locking unit once is
\begin{equation}
E_{S-SW1D} = NC_{g}(V_{g}-0)^{2} = NC_{g}V_{g}^2 .
\end{equation}
The capacitance of a MUX addressing gate is dominated by the area that is directly on top of the MESA, since the area on top of 500 $\mu$m thick substrate contributes negligibly and the area on top of the polyimide contributes at least 10 times smaller by design (as required by the selective gating method). In Figure \ref{CL-Schematics}a, the dominant contribution of gate capacitance is highlighted with red rectangles. The capacitance of each addressing gate is proportional to the number of channels that are effectively gated by it, and thus doubles for each higher level. To recharge  a charge-locking unit involves one addressing gate at each level being switched. If one addressing gate is switched once at each level, the total energy dissipated in MUX to select one charge-locking unit is equal to
\begin{equation}
  \begin{split}
E_{S-MUX} &= (2^{K-1} + 2^{K-2} .... + 1)C_{MUX}(2V_{g}-0)^{2}\\
        &= 4(N-1)C_{MUX}V_{g}^{2}
  \end{split},
\end{equation}
where $C_{MUX}$ is the capacitance contributed by each red rectangle. To sequentially recharge all the charge-locking units once, the total energy dissipated in analog MUX is thus equal to
\begin{equation}
E_{S-MUX1D} = NE_{S-MUX} = 4N(N-1)C_{MUX}V_{g}^{2}.
\label{MUX1D_1}
\end{equation}
Suppose each red rectangle has the same capacitance as the transistor of each charge-locking unit, i.e. $C_{MUX} = C_{g}$. Then, for a recharging frequency of $f_{c}$, the total power dissipated is equal to
\begin{equation}
  \begin{split}
P_{S-1D} &= (E_{S-MUX1D} + E_{S-SW1D})f_{c} \\
      &= (4N^2-3N)C_{g}V_{g}^2f_{c}
  \end{split}.
\end{equation}
The power dissipated to periodically recharge the 1D charge-locking circuitry grows quadratically with the total number of charge-locking units. The major contribution comes from switching the highest level MUX addressing gates.

In a similar way, we can calculate the power dissipated to periodically recharge the 2D charge-locking array with cross-bar addressing (Figure \ref{CL-Schematics}c). Each charge-locking unit is connected to two transistors in series, of which one can be switched on/off by shared row control line and the other can be switched on/off by shared column control line. As a result, each combination of row/column control line corresponds one charge-locking unit. $N$ column control lines and $M$ row control lines are able to control $N\times M$ charge-locking units. As each column control line is shared by $M$ transistors and each row control line is shared by $N$ transistors, it can be obtained that the energy dissipated to sequentially switch on/off the two transistors of each charge-locking unit once is
\begin{equation}
  \begin{split}
E_{S-SW2D} & = NM(NC_{g}+MC_{g})(V_{g}-0)^{2} \\
           & = NM(N+M)C_{g}V_{g}^{2}
  \end{split}.
\end{equation}

For controlling large numbers of charge-locking units, row and column control lines need to be multiplexed. The energy dissipated in MUX to select one row control line and one column control line once are $E_{row}$ and $E_{col}$ respectively, which can be expressed as
\begin{equation}
  \begin{split}
E_{S-row} = 4(M-1)C_{g}V_{g}^{2}\\
E_{S-col} = 4(N-1)C_{g}V_{g}^{2}
  \end{split}
\end{equation}
following the reasoning used to obtain $E_{S-MUX}$.
To sequentially recharge all the charge-locking units once, the total energy dissipated in two analog MUXs is thus equal to
\begin{equation}
  \begin{split}
E_{S-MUX2D} &= NM(E_{S-row}+E_{S-col}) \\
        &= 4NM(N+M-2)C_{g}V_{g}^{2}
\end{split}.
\end{equation}
If the periodic recharging is performed at a frequency of $f_{c}$, then the power dissipation is
\begin{equation}
  \begin{split}
P_{S-2D} &= (E_{S-MUX2D} + E_{S-SW2D})f_{c} \\
      &= 5NM(N+M-\frac{8}{5})C_{g}V_{g}^2f_{c}
  \end{split}.
\end{equation}
For a charge-locking array of equal size in both dimensions i.e. $ N = M $, the total number of charge-locking units is equal to $ N^2 $ and the total power dissipation $P_{S-2D} $ has a superlinear dependence (to the power of $\frac{3}{2}$) on the total number of charge-locking units. As opposed to the 1D charge-locking circuitry, the power dissipated in MUX no longer dominates and is only roughly 4 times as much as that is dissipated in switching transistors of charge-locking units. In other words, the conventional cross-bar addressing itself will lead to a superlinear growth in power dissipation regardless of the exact implementations of signal multiplexing circuitry. The ratio 4 is based on the assumption that $2V_{g}$ being applied to MUX addressing gates is required to route a DC signal of $V_{g}$. It can be smaller in principle, though that will not affect the superlinear growth of power dissipation. 

\section{Parallel refreshed charge-locking Approach}
In this section, we will present a different charge-locking approach, in which charge-locking units are recharged row-wise in a parallel manner, which greatly reduces the number of switchings required. The key difference between our approach and the conventional approach as described in Section \ref{PDA} is that the multiplexed outputs are not used to drive the transistors (switches) of charge-locking units, but rather are used for charging the holding capacitors, with consequent power reduction. Unlike the conventional approaches, there is completely no difference in implementing 1-dimensional and 2-dimensional charge-locking unit addressing with our approach.

\begin{figure*}[h]
\includegraphics[width=0.98 \linewidth]{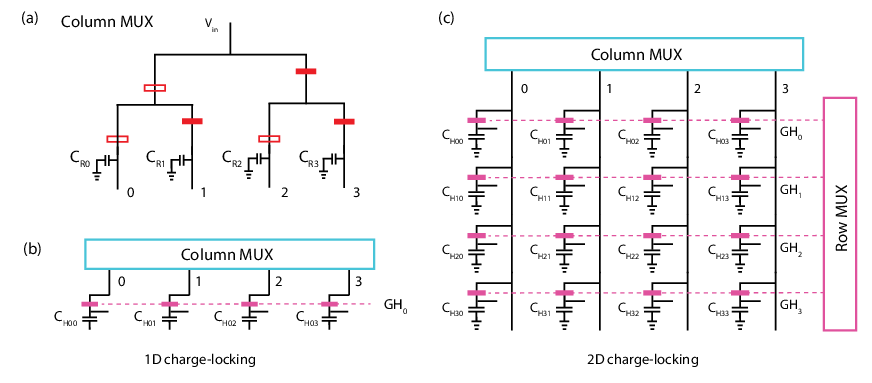}
\caption{(a) Column MUX (2-level base-2) for parallel recharging. The MUX itself is based on the MLSG method to route DC signals and each multiplexed output is connected to recharging capacitor $C_{Rj}$ (j = 0, 1, 2, 3). (b) 1-dimensional charge-locking circuitry: As opposed to the conventional charge-locking approaches, where the switches of charge-locking units are multiplexed and the input line for charging is shared, the multiplexed outputs of Column MUX are each connected to a holding capacitor $C_{H0j}$ (j =0, 1, 2, 3) and the switches are controlled by a shared control line. (c) To extend to two-dimensional charge-locking array, more rows of charge-locking units are added, where row-wise shared control lines are used for switching transistors. Row MUX is same as the Column MUX except that each multiplexed output is connected with a recharging capacitor.
\label{CL-Array}}
\end{figure*}

We will illustrate the basic charge-locking operation with a 2-level base-2 MUX, but the underlying principle can easily be extended to MUX of more levels. Figure \ref{CL-Array}a shows the column MUX used for performing parallel recharging, where each multiplexed output is connected to a recharging capacitor $C_{Rj}$, j = 0, 1, 2, 3. For 1D charge-locking, each multiplexed output is connected to a holding capacitor $C_{H0j}$, j = 0, 1, 2, 3, which can be simultaneously connected to column MUX for recharging and disconnected from column MUX for maintaining the static voltages, controlled by one shared control line e.g. $GH_{0}$ (Figure \ref{CL-Array}b). To extend to 2D charge-locking is essentially to add more rows of charge-locking units, where each row of charge-locking units share a control line to turn on/off transistors simultaneously (Figure \ref{CL-Array}c).

\begin{figure*}[h]
\includegraphics[width=0.98 \linewidth]{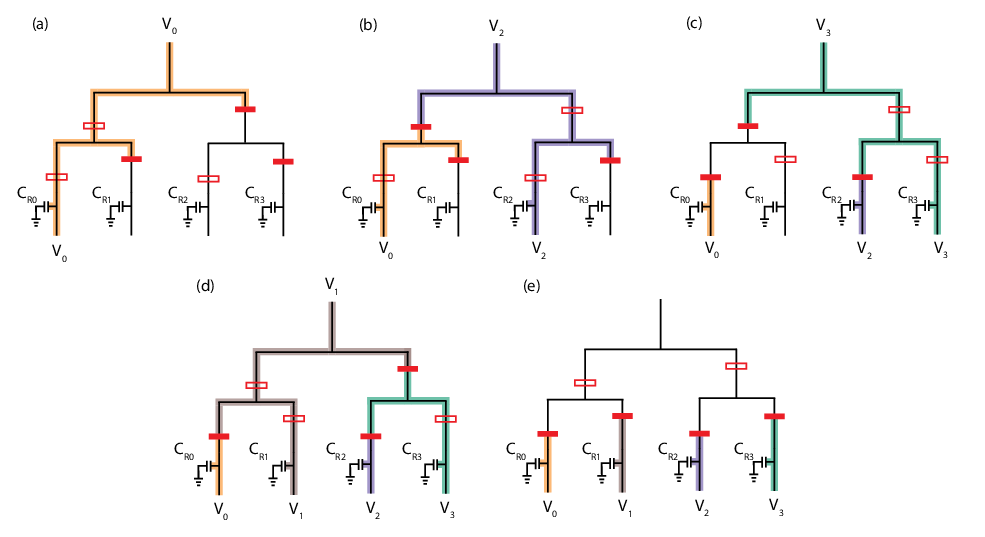}
\caption{Schematics showing the sequence to charge-lock 4 multiplexed output of Column MUX to hold different voltages, which is the core of our charge-locking approach and underpins both holding capacitor initialisation and parallel recharging process. By following this specific sequence of operating MUX addressing gates, the built-in switches of MUX allow simultaneously connecting one multiplexed output for charging while keeping the outputs that are already charged isolated. As a result, the total number of switchings is greatly reduced.
\label{CL-Process1}}
\end{figure*}

\begin{figure*}[h]
\includegraphics[width=0.98 \linewidth]{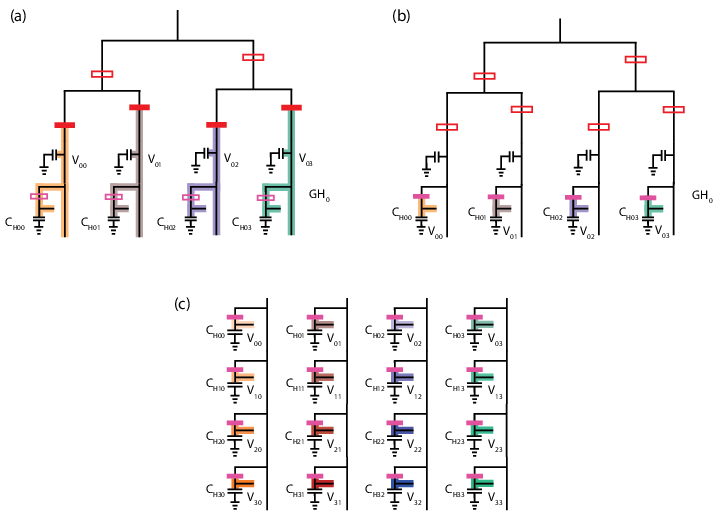}
\caption{(a) To initialise holding capacitor $C_{H0j} (j = 0,1,2,3)$, transistors are turned by by the shared control line $GH_{0}$. Then each holding capacitor can be charged to hold a different voltage following exactly the same sequence as shown in Figure \ref{CL-Process1}. (b) Transistors are turned off by the shared control line and all the holding capacitors are set into charge-locked state. (c) Each row of charge-locking units can be charged sequentially to hold different voltages by repeating steps shown in (a) and (b).
\label{CL-Process2}}
\end{figure*}

Figure \ref{CL-Process1} shows how each recharging capacitor of the column MUX is charge-locked to hold different static voltages, which is the core of our charge-locking approach (both initialising and recharging the holding capacitors). To initialise a specific row of charge-locking units to hold different voltages essentially follows the exactly same process except for connecting it to the column MUX beforehand, such that the same process will charge up both recharging capacitors and that row of holding capacitors (Figure \ref{CL-Process2}a). It can then be disconnected from the column MUX (Figure \ref{CL-Process2}b) without affecting the voltages being held while a different row of charge-locking units is being initialised. Charge-locking units can be initialised row by row to hold  different static voltages (Figure \ref{CL-Process2}c).

Key to the process shown in Figure \ref{CL-Process1} is to operate the addressing gates (i.e. built-in switches) of the column MUX in a  sequence that keeps the previously charged capacitors disconnected. As result, the number of switchings is greatly reduced as compared with the conventional approaches described in Section \ref{PDA}. The detailed procedure is as follows. We first activate addressing gate 1R/2R to select output 0 and set the input voltage to $V_{0}$, which will charge up capacitor $C_{R0}$ (Figure \ref{CL-Process1}a). We then activate addressing gate 1L and deactivate addressing gate 1R, which will disconnect capacitor $C_{R0}$ from the input and select output 2. Subsequently, the input voltage is set to $V_{2}$ to charge up capacitor $C_{R2}$, while the static voltage at output 0 stays unchanged (Figure \ref{CL-Process1}b). Thereafter, we activate addressing gate 2L and deactivate addressing gate 2R, to disconnect capacitor $C_{0}$ and $C_{2}$ from the input and select output 3. Up to here, output 0 and 1 are both in charge-locked state. Then the input is set to $V_{3}$ to charge capacitor $C_{R3}$ (Figure \ref{CL-Process1}c). Next we activate addressing gate 1R and deactivate addressing gate 1L, to disconnect capacitor $C_{R3}$ from the input and select output 1. The input is set to $V_{1}$ to charge capacitor $C_{R1}$ (Figure \ref{CL-Process1}d). Lastly, we activate addressing gate 2R to isolate capacitor $C_{R1}$ and $C_{R3}$. It is not essential to deactivate 1R here, the state in Figure \ref{CL-Process1}e simply shows four capacitors each being charged to hold a different static voltage. This charge-locking approach can be easily extended to base-2 MUX of more levels. Table \ref{MUX_16} shows the sequence to charge-lock 16 different static voltages with a 4-level base-2 MUX. Higher level gates, which have larger capacitance, are switched less frequently. The approach demonstrated in \citep{Puddy2015} requires one addressing gate of each level to be switched once for any charge-locking unit to be charged. Our less frequent switching of higher level addressing gates reduces the power dissipation.

\begin{table*}
  \begin{tabular}{|p{2cm}||p{2.5cm}||p{1cm}|p{1cm}|p{1cm}|p{1cm}|p{1cm}|p{1cm}|p{1cm}|p{1cm}|}
   \hline
   \hline
   Output No & Binary & 1L & 1R & 2L & 2R & 3L & 3R & 4L & 4R\\
   \hline
   0          & 0000                &OFF &ON  &OFF &ON  &OFF &ON  &OFF &ON\\
   8          & 1000                &ON  &OFF &OFF &ON  &OFF &ON  &OFF &ON\\
   12         & 1100                &ON  &OFF &ON  &OFF &OFF &ON  &OFF &ON\\
   4          & 0100                &OFF &ON  &ON  &OFF &OFF &ON  &OFF &ON\\
   6          & 0110                &OFF &ON  &ON  &OFF &ON  &OFF &OFF &ON\\
   14         & 1110                &ON  &OFF &ON  &OFF &ON  &OFF &OFF &ON\\
   10         & 1010                &ON  &OFF &OFF &ON  &ON  &OFF &OFF &ON\\
   2          & 0010                &OFF &ON  &OFF &ON  &ON  &OFF &OFF &ON\\
   3          & 0011                &OFF &ON  &OFF &ON  &ON  &OFF &ON  &OFF\\
   11         & 1011                &ON  &OFF &OFF &ON  &ON  &OFF &ON  &OFF\\
   15         & 1111                &ON  &OFF &ON  &OFF &ON  &OFF &ON  &OFF\\
   7          & 0111                &OFF &ON  &ON  &OFF &ON  &OFF &ON  &OFF\\
   5          & 0101                &OFF &ON  &ON  &OFF &OFF &ON  &ON  &OFF\\
   13         & 1101                &ON  &OFF &ON  &OFF &OFF &ON  &ON  &OFF\\
   9          & 1001                &ON  &OFF &OFF &ON  &OFF &ON  &ON  &OFF\\
   1          & 0001                &OFF &ON  &OFF &ON  &OFF &ON  &ON  &OFF\\
   \hline
 \end{tabular}
 \caption{The sequence to charge-lock static voltages with a 4-level base-2 MUX. In the table, the activated state of addressing gate is labelled as ON and deactivated state of addressing gate is labelled as OFF. The relationship between the selected output and the state of addressing gates is given in binary representation. At each level, an activated L addressing gate corresponds to 1 and an activated R addressing gate corresponds 0. For example, output 11 is selected by activating 1L/2R/3L/4L, which is 1011 in binary and 11 in decimal. \label{MUX_16}}
\end{table*}

\begin{figure*}[h]
\includegraphics[width=0.98 \linewidth]{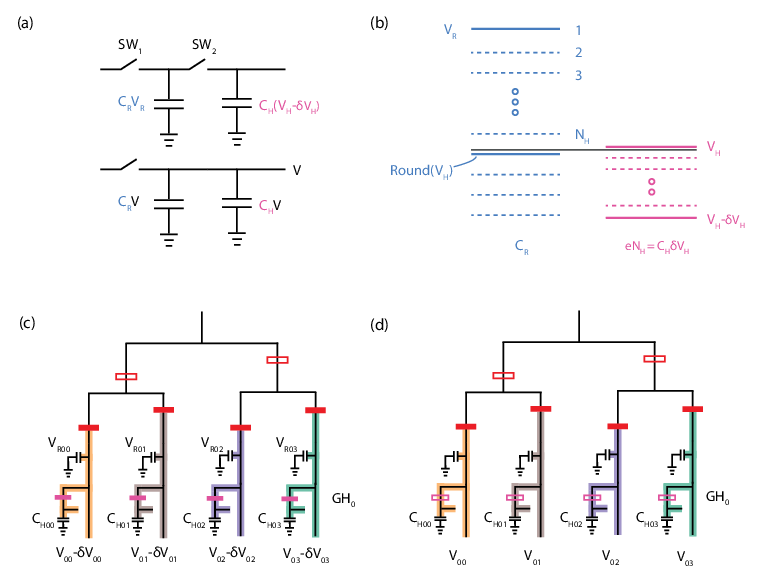}
\caption{(a) Recharging capacitor $C_{R}$ and holding capacitor $C_{H}$ are with two different voltages $V_{R}$ and $V_{H} - \delta V_{H}$. Once they are connected with each other, charge redistributes among them, and an equilibrium voltage $V$ is reached. This charge redistributing process can be employed to compensate the voltage drift $\delta V_{H}$ in holding capacitor if $C_{R}$ is set to the right voltage. (b) The voltage drift in $C_{H}$ corresponds to a number of $N_{H} = \frac{C_{H} \delta V_{H}}{e}$ electrons to be compensated. To ensure $N_{H}$ electrons flowing into $C_{H}$ from $C_{R}$ during charge redistribution, $C_{R}$ should be prepared to a voltage level $V_{R}$,  which is equal to $Round(V_{H})$ + $N_{H}e/C_{R}$. $Round(V_{H})$ accounts for the charge-discreteness and is the highest voltage level of $C_{R}$ below $V_{H}$. $e/C_{R}$ is the voltage required to add each more electron to $C_{R}$. (c) All the holding capacitors $C_{H0j} (j =0, 1, 2, 3)$ are kept at charge-locked state, while the recharging capacitors are prepared to hold the right voltage for recharging. (d) The transistors are turned on by the shared control line $GH_{0}$ and charge redistributes among recharging capacitors and holding capactiors, such that holding capacitors are recharged in a parallel way.
\label{CL-Recharging}}
\end{figure*}

Next, we will describe the procedure to perform periodic recharging. The basic idea is illustrated in Figure \ref{CL-Recharging}a. Static voltages of $V_{R}$ and $V_{H} - \delta V_{H}$ are held by recharging capacitor $C_{R}$ and holding $C_{H}$ respectively. When they are connected to each other, the charge redistributes among them, and an equilibrium voltage is reached and is equal to
\begin{equation}
V = \frac{V_{R}C_{R}+(V_{H}-\delta V_{H}) C_{H}}{C_{R}+C_{H}}.
\end{equation}
If $\delta V_{H}$ is the voltage to be compensated and to restore the original voltage requires $V = V_{H}$, then
\begin{equation}
V_{R} = V_{H} + \frac {C_{H}\delta V_{H}}{C_{R}}.
\end{equation}
It might appear that the recharging capacitor $C_{R}$ has to be at least as big as the holding capacitor $C_{H}$ to precisely recharge the holding capacitor, but that is not necessary. To account for the effect of charge discreteness (Figure \ref{CL-Recharging}b), we introduce a new parameter $N_{H}$, which is the number of electrons leaked from the holding capacitor $C_{H}$ and can be expressed as
\begin{equation}
N_{H} = \frac{C_{H}\delta V_{H}}{e}.
\end{equation}
To ensure $N_{H}$ flowing back to $C_{H}$ during recharging, the recharging capacitor $C_{R}$ should be charged up to

\begin{equation}
V_{R} = Round (V_{H}) + N_{H}\frac{e}{C_{R}},
\label{Eq-RC}
\end{equation}
where $Round (V_{H})$ is the voltage to be restored rounded to a coarse resolution $\frac{e}{C_{R}}$. For example, a voltage of 1411.1~$\mu$V rounded to a resolution of 10~$\mu$V will be 1410~$\mu$V. As discussed above, to reach a voltage resolution of 1~$\mu$V at 100~mK, the holding capacitor $C_{H}$ has to be at least 1.4~pF and is set by the uncertainty introduced by thermal noise $\sqrt{\frac{KT}{C_{H}}}$. The resolution set by charge discreteness is $e/C_{H} \sim$ 0.1 $\mu$V. If a recharging capacitor $C_{R} < 0.01 C_{H}$ is used for recharging, the voltage resolution of the isolated $C_{R}$ is determined by charge discreteness and $e/C_{R}$ will be larger than 10 $\mu$V. As will be shown later, the choice of $C_{R}$ is relatively flexible for charge-locking circuitry with 1D addressing. For a 2D charge-locking array, a recharging capacitor much smaller than the holding capacitor reduces power dissipation.

Figure \ref{CL-Recharging}c shows each recharging capacitor $C_{R0j}$ is charged to hold a static voltage $V_{R0j}$ and prepared for recharging holding capacitors. Then all the transistors are turned on by the shared control line $GH_{0}$ and charge is redistributed to compensate for the voltage drift of each holding capacitor (Figure \ref{CL-Recharging}d). In practice, $V_{R0j}$ here accounts for not only the voltage drift $\delta V_{0j}$ arising from charge leakage (as determined by Equation \ref{Eq-RC}) but also the systematic offset induced by charge injection from the transistor gate when the transistor is turned off, for each charge-locked voltage, which can be calibrated from the current at that specific charge-locked voltage \citep{Pauka2019a,Xu2020,Puddy2015}. The systematic offset induced by charge injection from the transistor gate depends on the relative size of the transistor gate and the holding capacitor, and for large transistor sizes can even be comparable to the voltage drift (\citep{Xu2020}). If $C_{H}$ and $C_{g}$ are of size 10~$\mu$m $\times$ 14~$\mu$m and 10~nm $\times$ 14~nm respectively, $C_{H}$ has a resolution of $\sim$ 0.1 $\mu$V due to charge discreteness and $C_{g}$ correspondingly has a resolution of $\sim$ 100~mV. For $V_{g} \simeq$ 1V applied to the transistor gate, about 10 electrons are confined in the transistor channel. If 5 electrons are being injected to $C_{H}$, 0.5 $\mu$V systematic offset is induced, which is smaller than the uncertainty induced by thermal noise.

Next, we will begin the analysis of the power dissipated in periodic recharging for both 1D and 2D charge-locking. For 1D charge-locking, suppose there are in total N charge-locking units, which are controlled by a K-level base-2 analog MUX (i.e. $N= 2^{K}$). As discussed in Section \ref{PDA}, the total power dissipation still consists of two major contributions, it is easy to obtain that the energy dissipated to switch on/off transistors once is
\begin{equation}
E_{P-SW1D} = NC_{g}(V_{g}-0)^{2} = NC_{g}V_{g}^2 .
\end{equation}
To obtain the energy dissipated in operating the analog MUX, we observe that each i-th level addressing gate is switched on/off every $2^{i}$ outputs in the sequence to be set to charge-locked state (see Table \ref{MUX_16}), of which the capacitance is $2^{i-1}C_{g}$. To prepare all the recharging capacitors, two addressing gates at each level will in total dissipate
\begin{equation}
  \begin{split}
E_{P-MUX1D}     &= 2\times K \times \frac{2^{K}}{2^{i}} \times 2^{i-1} \times C_{g}V_{g}^{2} \\
      &= KNC_{g}V_{g}^{2}
\end{split}.
\end{equation}
The voltage applied to the addressing gates is V\textsubscript{g} instead of 2V\textsubscript{g}, leading to further power reduction. The factor K accounts for the total K-level addressing gates. For a recharging frequency of $f_{c}$, the total power dissipated will be
\begin{equation}
  \begin{split}
P_{P-1D}&= (E_{P-SW1D}+E_{P-MUX1D})f_{c} \\
      & = (1+K)NC_{g}V_{g}^{2}f_{c} .
  \end{split}
\end{equation}
Equivalently, the total power dissipated can be expressed in the total number of charge-locking units $N$
\begin{equation}
P_{P-1D}=(1+ \log_2 N)NC_{g}V_{g}^{2}f_{c} .
\end{equation}
$P_{P-1D}$ is very close to linearly growing with the total number of charge-locking units $N$, since ${\log_2(N+1)}- {\log_2 (N)} \simeq 0$ for large $N$.

Similarly, for a 2D charge-locking array that consists of M rows and N columns, i.e. of $N \times M$ charge-locking units in total. The energy dissipated in turning on/off the transistors of every charge-locking unit once is
\begin{equation}
E_{P-SW2D} = M \times E_{P-SW1D} = NMC_{g}V_{g}^2 .
\end{equation}
Then, the energy dissipated in column MUX is
\begin{equation}
  \begin{split}
E_{P-col} & = M \times E_{P-MUX1D} \\
          & = NMKC_{g}V_{g}^2 .
  \end{split}
\end{equation}
As for the energy dissipated in row MUX, it can be obtained following the same reasoning to Equation \ref{MUX1D_1}
\begin{equation}
E_{P-row} = 4M(M-1)C_{g}V_{g}^{2} .
\end{equation}
In addition to the above three contributions, there is a further contribution that is associated with preparing each recharging capacitor to the right voltage for recharging different holding capacitors, since the voltage held by each charge-locking unit varies along each column. If the RMS (root mean square) value of the voltage variation along each column is $\delta V_{R}$, then the total energy dissipated here is
\begin{equation}
E_{RC} = NMC_{R}\delta V_{R}^2 .
\label{Eq-RC2}
\end{equation}
Here, we have assumed the $C_{R}$ is chosen to be sufficiently large, such that the voltage swing involved in preparing $C_{R}$ for recharging different holding capacitors along the column is still dominated by the variation between each voltage being held.

For a recharging frequency of $f_{c}$ and array of equal size in both dimensions, the total power dissipation is
\begin{equation}
  \begin{split}
P_{P-2D}& = (E_{P-SW2D}+E_{P-col}+E_{P-row}+E_{RC})f_{c} \\
      & = N^{2} (1+ \log_2 N + 4\frac{N-1}{N} + Q)C_{g}V_{g}^2f_{c} \\
      & \simeq N^{2} (5 + \log_2 N + Q)C_{g}V_{g}^2f_{c} .
  \end{split}
\end{equation}
where $Q = \frac{C_{R}\delta V_{R}^2}{C_{g}V_{g}^2 }$ and can be minimised by choosing the smallest $C_{R}$ possible. The choice of $C_{R}$ is largely determined by $\delta V_{R}$. In an experiment on shuttling spin over a 9 QD array, the RMS value of the plunger gate voltage variation was calculated to be 47 mV (based on the Supplementary Table 1 in \citep{Mills2019}). Below we will calculate a upper bound for Q using $\delta V_{R} = 100$ mV and a lower bound for Q using $\delta V_{R} = 10$ mV. Given the currently achieved variance of 47 mV and the demonstration of low disorder double QDs \citep{Borselli2015},  $\delta V_{R} = 10$ mV is very likely to be achieved with further optimisation. 

In the case of $\delta V_{H} = 100$ mV $\simeq \frac{1}{10}V_{g}$, the recharging capacitor $C_{R}$ can be as small as $\frac{1}{1000} C_{H}$, such that a voltage drift of 10 $\mu$V in $C_{H}$ corresponds roughly to 10 mV in $C_{R}$ (see Equation \ref{Eq-RC}), which is still an order of magnitude smaller than $\delta V_{H}$ and ensures Equation \ref{Eq-RC2} still being valid. Suppose $C_{g}$ is of the size 14~nm $\times$ 10~nm, which is $\frac{1}{10^{6}}C_{H} = \frac{1}{10^{3}}C_{R}$, then Q is calculated to be 10.

Similarly, in the case of $\delta V_{H} = 10$ mV $\simeq \frac{1}{100}V_{g}$, the recharging capacitor $C_{R}$ can be as small as $\frac{1}{100} C_{H}$, such that a voltage drift of 10 $\mu$V in $C_{H}$ corresponds roughly to 1mV in $C_{R}$, which makes sure Equation \ref{Eq-RC2} is still valid. Suppose $C_{g}$ is of the size 14~nm $\times$ 10~nm, which is $\frac{1}{10^{6}}C_{H} = \frac{1}{10^{4}}C_{R}$, then Q is calculated to be 1. From the value of Q, we conclude that this additional power dissipation involved in preparing recharging capacitor is comparable to other two contributions. $P_{P-2D}$ is also very close to linear growth in power dissipation as the array size scales up. Unlike conventional approaches, improvements in uniformity lead to reductions in power dissipation.

\section{Discussion and Conclusion}
Below we will evaluate the power dissipation required to maintain $2^{14} \times 2^{14} \simeq 2.6 \times 10^{8}$ charge-locking units with both serially and parallel refreshed 2D charge-locking solutions, and benchmark them with respect to the cooling power of large dilution fridge available at 100mK, which is around 1~mW. It is generally accepted that order of $10^{8}$ qubits are required for running error correction code to reach fault-tolerant quantum computation \citep{Veldhorst2017,Vandersypen2017}. The detailed calculation is shown in SI. For serially refreshed charge-locking solution with conventional 2D cross-bar addressing approach, the power dissipated to maintain $2.6 \times 10^{8}$ charge-locking units is around 61.5~mW per kHz refreshing rate. In contrast, the parallel refreshed charge-locking solution developed in this paper only dissipates 11~$\mu$W per kHz refreshing rate (assuming variation $\delta V_{R} \simeq 100$~mV and $Q \simeq 10$), which is more than 5000 times reduction in power dissipation. If the variation $\delta V_{R}$ is improved to be around 10~mV, the power dissipation can be further reduced to be around 7.5~ $\mu$W per kHz refreshing rate. As a comparison, an alternative solution to allow local control electronics residing closely with qubits is to operate them at 4K, of which the cooling power is around 1W and is thus only with 1000 fold increase. In other words, a parallel refreshed charge-locking array implemented at 100mK can be larger in size as compared with a serially refreshed charge-locking array implemented at 4K, solely from a power dissipation perspective. 

Another more subtle advantage of our approach is related to the recharging frequency $f_{c}$. The minimum recharge frequency is determined by the leakage current and allowed static voltage drift. On the other hand, the time $T$ required to recharge all the charge-locking units sets the upper limit for this recharge frequency, which in turn limits the total number of charge-locked signals that can be maintained. Similar reduction in recharging time is achieved with parallel refreshed charge-locking approach as compared with the conventional approaches (See SI for detailed analysis). For any charge-locking array with the size smaller than $10^{12}$, it can be shown that the parallel refreshed charge-locking array show a constant $\frac{C_{H}}{C_{R}}$ fold reduction in total recharging time, essentially increasing the upper limit for charge-locking size by the same ratio. 

Base-4 signal multiplexing based on MSLG method (Figure \ref{Base-4}b) offers further reduction in the power dissipation compared with base-2 multiplex scheme. For the same number of control lines, a base-4 MUX scheme allows as many multiplexed outputs as base-2 signal multiplexing. The power reduction is obtained by replacing the $\log_2 N$ term in $P_{P-2D}$ with a $\log_4 N$ term. As a result, it will lead to another 35 $\%$ reduction in power dissipation as compared with the base-2 addressing. This parallel refreshed charge-locking scheme can be generalised for any MLSG based MUX with arbitrary heterogeneous base stacking (See SI for two examples). With different base stacking, it adds another layer of flexibility to sub-divide the entire array, allowing different combinations of charge-locking units to be simultaneously selected. As a result, with additional local capacitors integrated, column-wise DC pulsing can be easily performed on specific rows as required by qubit operation.  

To conclude, our parallel refreshed charge-locking approach yields significantly lower power dissipation than conventional approaches. Whereas conventional approaches all show superlinear dependence, in our system the power dissipation grows approximately linearly with the total number charge-locking units. The power dissipated to maintain $2^{14} \times 2^{14} = 2.6 \times 10^{8}$ charge-locking units is as low as 11 ~$\mu$W per kHz recharging frequency. The power dissipation can be further reduced with the continuing improvement in QD uniformity. For a voltage variation of $\delta V_{H} \simeq 10$~mV, it is expected to reach 7.5 ~$\mu$W per kHz recharging frequency. The total recharging time is also largely reduced with parallel recharging, which means one critical upper limit for the charge-locking array size is essentially increased. This charge-locking array requires minimal number of connections from higher temperature. In principle, 58 lines fed from higher temperature, 28 lines for the addressing gates of each MUX and 1 for the input of each MUX, will suffice to operate 100 million charge-locking units. In practice, there is another limit imposed by digital electronics (a few GHz switching rate). To resolve this problem, multiple independent inputs for Column MUX (as the structure used in \citep{Ward2013}) can be used to reduce the required switching rate to a few GHz, which however will not change the overall power dissipation for the same array size. (See details in SI section A)

\bibliography{MUX}

\begin{thebibliography}{32}
\expandafter\ifx\csname natexlab\endcsname\relax\def\natexlab#1{#1}\fi
\expandafter\ifx\csname bibnamefont\endcsname\relax
  \def\bibnamefont#1{#1}\fi
\expandafter\ifx\csname bibfnamefont\endcsname\relax
  \def\bibfnamefont#1{#1}\fi
\expandafter\ifx\csname citenamefont\endcsname\relax
  \def\citenamefont#1{#1}\fi
\expandafter\ifx\csname url\endcsname\relax
  \def\url#1{\texttt{#1}}\fi
\expandafter\ifx\csname urlprefix\endcsname\relax\def\urlprefix{URL }\fi
\providecommand{\bibinfo}[2]{#2}
\providecommand{\eprint}[2][]{\url{#2}}

\bibitem[{\citenamefont{Almudever et~al.}(2017)\citenamefont{Almudever, Lao,
  Fu, Khammassi, Ashraf, Iorga, Varsamopoulos, Eichler, Wallraff, Geck
  et~al.}}]{Almudever2017}
\bibinfo{author}{\bibfnamefont{C.~G.} \bibnamefont{Almudever}},
  \bibinfo{author}{\bibfnamefont{L.}~\bibnamefont{Lao}},
  \bibinfo{author}{\bibfnamefont{X.}~\bibnamefont{Fu}},
  \bibinfo{author}{\bibfnamefont{N.}~\bibnamefont{Khammassi}},
  \bibinfo{author}{\bibfnamefont{I.}~\bibnamefont{Ashraf}},
  \bibinfo{author}{\bibfnamefont{D.}~\bibnamefont{Iorga}},
  \bibinfo{author}{\bibfnamefont{S.}~\bibnamefont{Varsamopoulos}},
  \bibinfo{author}{\bibfnamefont{C.}~\bibnamefont{Eichler}},
  \bibinfo{author}{\bibfnamefont{A.}~\bibnamefont{Wallraff}},
  \bibinfo{author}{\bibfnamefont{L.}~\bibnamefont{Geck}}, \bibnamefont{et~al.},
  in \emph{\bibinfo{booktitle}{Design, Automation {\&} Test in Europe
  Conference {\&} Exhibition ({DATE}), 2017}} (\bibinfo{publisher}{{IEEE}},
  \bibinfo{year}{2017}).

\bibitem[{\citenamefont{Franke et~al.}(2019)\citenamefont{Franke, Clarke,
  Vandersypen, and Veldhorst}}]{Franke2019}
\bibinfo{author}{\bibfnamefont{D.}~\bibnamefont{Franke}},
  \bibinfo{author}{\bibfnamefont{J.}~\bibnamefont{Clarke}},
  \bibinfo{author}{\bibfnamefont{L.}~\bibnamefont{Vandersypen}},
  \bibnamefont{and}
  \bibinfo{author}{\bibfnamefont{M.}~\bibnamefont{Veldhorst}},
  \bibinfo{journal}{Microprocessors and Microsystems}
  \textbf{\bibinfo{volume}{67}}, \bibinfo{pages}{1} (\bibinfo{year}{2019}).

\bibitem[{\citenamefont{Vandersypen et~al.}(2017)\citenamefont{Vandersypen,
  Bluhm, Clarke, Dzurak, Ishihara, Morello, Reilly, Schreiber, and
  Veldhorst}}]{Vandersypen2017}
\bibinfo{author}{\bibfnamefont{L.~M.~K.} \bibnamefont{Vandersypen}},
  \bibinfo{author}{\bibfnamefont{H.}~\bibnamefont{Bluhm}},
  \bibinfo{author}{\bibfnamefont{J.~S.} \bibnamefont{Clarke}},
  \bibinfo{author}{\bibfnamefont{A.~S.} \bibnamefont{Dzurak}},
  \bibinfo{author}{\bibfnamefont{R.}~\bibnamefont{Ishihara}},
  \bibinfo{author}{\bibfnamefont{A.}~\bibnamefont{Morello}},
  \bibinfo{author}{\bibfnamefont{D.~J.} \bibnamefont{Reilly}},
  \bibinfo{author}{\bibfnamefont{L.~R.} \bibnamefont{Schreiber}},
  \bibnamefont{and}
  \bibinfo{author}{\bibfnamefont{M.}~\bibnamefont{Veldhorst}},
  \bibinfo{journal}{npj Quantum Information} \textbf{\bibinfo{volume}{3}}
  (\bibinfo{year}{2017}).

\bibitem[{\citenamefont{Hornibrook et~al.}(2015)\citenamefont{Hornibrook,
  Colless, Lamb, Pauka, Lu, Gossard, Watson, Gardner, Fallahi, Manfra
  et~al.}}]{Hornibrook2015}
\bibinfo{author}{\bibfnamefont{J.}~\bibnamefont{Hornibrook}},
  \bibinfo{author}{\bibfnamefont{J.}~\bibnamefont{Colless}},
  \bibinfo{author}{\bibfnamefont{I.~C.} \bibnamefont{Lamb}},
  \bibinfo{author}{\bibfnamefont{S.}~\bibnamefont{Pauka}},
  \bibinfo{author}{\bibfnamefont{H.}~\bibnamefont{Lu}},
  \bibinfo{author}{\bibfnamefont{A.}~\bibnamefont{Gossard}},
  \bibinfo{author}{\bibfnamefont{J.}~\bibnamefont{Watson}},
  \bibinfo{author}{\bibfnamefont{G.}~\bibnamefont{Gardner}},
  \bibinfo{author}{\bibfnamefont{S.}~\bibnamefont{Fallahi}},
  \bibinfo{author}{\bibfnamefont{M.}~\bibnamefont{Manfra}},
  \bibnamefont{et~al.}, \bibinfo{journal}{Physical Review Applied}
  \textbf{\bibinfo{volume}{3}} (\bibinfo{year}{2015}).

\bibitem[{\citenamefont{Reilly}(2015)}]{Reilly2015}
\bibinfo{author}{\bibfnamefont{D.~J.} \bibnamefont{Reilly}},
  \bibinfo{journal}{npj Quantum Information} \textbf{\bibinfo{volume}{1}}
  (\bibinfo{year}{2015}).

\bibitem[{\citenamefont{Petta}(2005)}]{Petta2005}
\bibinfo{author}{\bibfnamefont{J.~R.} \bibnamefont{Petta}},
  \bibinfo{journal}{Science} \textbf{\bibinfo{volume}{309}},
  \bibinfo{pages}{2180} (\bibinfo{year}{2005}).

\bibitem[{\citenamefont{Veldhorst et~al.}(2014)\citenamefont{Veldhorst, Hwang,
  Yang, Leenstra, de~Ronde, Dehollain, Muhonen, Hudson, Itoh, Morello
  et~al.}}]{Veldhorst2014}
\bibinfo{author}{\bibfnamefont{M.}~\bibnamefont{Veldhorst}},
  \bibinfo{author}{\bibfnamefont{J.~C.~C.} \bibnamefont{Hwang}},
  \bibinfo{author}{\bibfnamefont{C.~H.} \bibnamefont{Yang}},
  \bibinfo{author}{\bibfnamefont{A.~W.} \bibnamefont{Leenstra}},
  \bibinfo{author}{\bibfnamefont{B.}~\bibnamefont{de~Ronde}},
  \bibinfo{author}{\bibfnamefont{J.~P.} \bibnamefont{Dehollain}},
  \bibinfo{author}{\bibfnamefont{J.~T.} \bibnamefont{Muhonen}},
  \bibinfo{author}{\bibfnamefont{F.~E.} \bibnamefont{Hudson}},
  \bibinfo{author}{\bibfnamefont{K.~M.} \bibnamefont{Itoh}},
  \bibinfo{author}{\bibfnamefont{A.}~\bibnamefont{Morello}},
  \bibnamefont{et~al.}, \bibinfo{journal}{Nature Nanotechnology}
  \textbf{\bibinfo{volume}{9}}, \bibinfo{pages}{981} (\bibinfo{year}{2014}).

\bibitem[{\citenamefont{Muhonen et~al.}(2014)\citenamefont{Muhonen, Dehollain,
  Laucht, Hudson, Kalra, Sekiguchi, Itoh, Jamieson, McCallum, Dzurak
  et~al.}}]{Muhonen2014}
\bibinfo{author}{\bibfnamefont{J.~T.} \bibnamefont{Muhonen}},
  \bibinfo{author}{\bibfnamefont{J.~P.} \bibnamefont{Dehollain}},
  \bibinfo{author}{\bibfnamefont{A.}~\bibnamefont{Laucht}},
  \bibinfo{author}{\bibfnamefont{F.~E.} \bibnamefont{Hudson}},
  \bibinfo{author}{\bibfnamefont{R.}~\bibnamefont{Kalra}},
  \bibinfo{author}{\bibfnamefont{T.}~\bibnamefont{Sekiguchi}},
  \bibinfo{author}{\bibfnamefont{K.~M.} \bibnamefont{Itoh}},
  \bibinfo{author}{\bibfnamefont{D.~N.} \bibnamefont{Jamieson}},
  \bibinfo{author}{\bibfnamefont{J.~C.} \bibnamefont{McCallum}},
  \bibinfo{author}{\bibfnamefont{A.~S.} \bibnamefont{Dzurak}},
  \bibnamefont{et~al.}, \bibinfo{journal}{Nature Nanotechnology}
  \textbf{\bibinfo{volume}{9}}, \bibinfo{pages}{986} (\bibinfo{year}{2014}).

\bibitem[{\citenamefont{Veldhorst et~al.}(2015)\citenamefont{Veldhorst, Yang,
  Hwang, Huang, Dehollain, Muhonen, Simmons, Laucht, Hudson, Itoh
  et~al.}}]{Veldhorst2015}
\bibinfo{author}{\bibfnamefont{M.}~\bibnamefont{Veldhorst}},
  \bibinfo{author}{\bibfnamefont{C.~H.} \bibnamefont{Yang}},
  \bibinfo{author}{\bibfnamefont{J.~C.~C.} \bibnamefont{Hwang}},
  \bibinfo{author}{\bibfnamefont{W.}~\bibnamefont{Huang}},
  \bibinfo{author}{\bibfnamefont{J.~P.} \bibnamefont{Dehollain}},
  \bibinfo{author}{\bibfnamefont{J.~T.} \bibnamefont{Muhonen}},
  \bibinfo{author}{\bibfnamefont{S.}~\bibnamefont{Simmons}},
  \bibinfo{author}{\bibfnamefont{A.}~\bibnamefont{Laucht}},
  \bibinfo{author}{\bibfnamefont{F.~E.} \bibnamefont{Hudson}},
  \bibinfo{author}{\bibfnamefont{K.~M.} \bibnamefont{Itoh}},
  \bibnamefont{et~al.}, \bibinfo{journal}{Nature}
  \textbf{\bibinfo{volume}{526}}, \bibinfo{pages}{410} (\bibinfo{year}{2015}).

\bibitem[{\citenamefont{Watson et~al.}(2018)\citenamefont{Watson, Philips,
  Kawakami, Ward, Scarlino, Veldhorst, Savage, Lagally, Friesen, Coppersmith
  et~al.}}]{Watson2018}
\bibinfo{author}{\bibfnamefont{T.~F.} \bibnamefont{Watson}},
  \bibinfo{author}{\bibfnamefont{S.~G.~J.} \bibnamefont{Philips}},
  \bibinfo{author}{\bibfnamefont{E.}~\bibnamefont{Kawakami}},
  \bibinfo{author}{\bibfnamefont{D.~R.} \bibnamefont{Ward}},
  \bibinfo{author}{\bibfnamefont{P.}~\bibnamefont{Scarlino}},
  \bibinfo{author}{\bibfnamefont{M.}~\bibnamefont{Veldhorst}},
  \bibinfo{author}{\bibfnamefont{D.~E.} \bibnamefont{Savage}},
  \bibinfo{author}{\bibfnamefont{M.~G.} \bibnamefont{Lagally}},
  \bibinfo{author}{\bibfnamefont{M.}~\bibnamefont{Friesen}},
  \bibinfo{author}{\bibfnamefont{S.~N.} \bibnamefont{Coppersmith}},
  \bibnamefont{et~al.}, \bibinfo{journal}{Nature}
  \textbf{\bibinfo{volume}{555}}, \bibinfo{pages}{633} (\bibinfo{year}{2018}).

\bibitem[{\citenamefont{Zajac et~al.}(2017)\citenamefont{Zajac, Sigillito,
  Russ, Borjans, Taylor, Burkard, and Petta}}]{Zajac2017}
\bibinfo{author}{\bibfnamefont{D.~M.} \bibnamefont{Zajac}},
  \bibinfo{author}{\bibfnamefont{A.~J.} \bibnamefont{Sigillito}},
  \bibinfo{author}{\bibfnamefont{M.}~\bibnamefont{Russ}},
  \bibinfo{author}{\bibfnamefont{F.}~\bibnamefont{Borjans}},
  \bibinfo{author}{\bibfnamefont{J.~M.} \bibnamefont{Taylor}},
  \bibinfo{author}{\bibfnamefont{G.}~\bibnamefont{Burkard}}, \bibnamefont{and}
  \bibinfo{author}{\bibfnamefont{J.~R.} \bibnamefont{Petta}},
  \bibinfo{journal}{Science} \textbf{\bibinfo{volume}{359}},
  \bibinfo{pages}{439} (\bibinfo{year}{2017}).

\bibitem[{\citenamefont{Huang et~al.}(2019)\citenamefont{Huang, Yang, Chan,
  Tanttu, Hensen, Leon, Fogarty, Hwang, Hudson, Itoh et~al.}}]{Huang2019}
\bibinfo{author}{\bibfnamefont{W.}~\bibnamefont{Huang}},
  \bibinfo{author}{\bibfnamefont{C.~H.} \bibnamefont{Yang}},
  \bibinfo{author}{\bibfnamefont{K.~W.} \bibnamefont{Chan}},
  \bibinfo{author}{\bibfnamefont{T.}~\bibnamefont{Tanttu}},
  \bibinfo{author}{\bibfnamefont{B.}~\bibnamefont{Hensen}},
  \bibinfo{author}{\bibfnamefont{R.~C.~C.} \bibnamefont{Leon}},
  \bibinfo{author}{\bibfnamefont{M.~A.} \bibnamefont{Fogarty}},
  \bibinfo{author}{\bibfnamefont{J.~C.~C.} \bibnamefont{Hwang}},
  \bibinfo{author}{\bibfnamefont{F.~E.} \bibnamefont{Hudson}},
  \bibinfo{author}{\bibfnamefont{K.~M.} \bibnamefont{Itoh}},
  \bibnamefont{et~al.}, \bibinfo{journal}{Nature}
  \textbf{\bibinfo{volume}{569}}, \bibinfo{pages}{532} (\bibinfo{year}{2019}).

\bibitem[{\citenamefont{Yang et~al.}(2020)\citenamefont{Yang, Leon, Hwang,
  Saraiva, Tanttu, Huang, Lemyre, Chan, Tan, Hudson et~al.}}]{Yang2020}
\bibinfo{author}{\bibfnamefont{C.~H.} \bibnamefont{Yang}},
  \bibinfo{author}{\bibfnamefont{R.~C.~C.} \bibnamefont{Leon}},
  \bibinfo{author}{\bibfnamefont{J.~C.~C.} \bibnamefont{Hwang}},
  \bibinfo{author}{\bibfnamefont{A.}~\bibnamefont{Saraiva}},
  \bibinfo{author}{\bibfnamefont{T.}~\bibnamefont{Tanttu}},
  \bibinfo{author}{\bibfnamefont{W.}~\bibnamefont{Huang}},
  \bibinfo{author}{\bibfnamefont{J.~C.} \bibnamefont{Lemyre}},
  \bibinfo{author}{\bibfnamefont{K.~W.} \bibnamefont{Chan}},
  \bibinfo{author}{\bibfnamefont{K.~Y.} \bibnamefont{Tan}},
  \bibinfo{author}{\bibfnamefont{F.~E.} \bibnamefont{Hudson}},
  \bibnamefont{et~al.}, \bibinfo{journal}{Nature}
  \textbf{\bibinfo{volume}{580}}, \bibinfo{pages}{350} (\bibinfo{year}{2020}).

\bibitem[{\citenamefont{Petit et~al.}(2020)\citenamefont{Petit, Eenink, Russ,
  Lawrie, Hendrickx, Philips, Clarke, Vandersypen, and Veldhorst}}]{Petit2020}
\bibinfo{author}{\bibfnamefont{L.}~\bibnamefont{Petit}},
  \bibinfo{author}{\bibfnamefont{H.~G.~J.} \bibnamefont{Eenink}},
  \bibinfo{author}{\bibfnamefont{M.}~\bibnamefont{Russ}},
  \bibinfo{author}{\bibfnamefont{W.~I.~L.} \bibnamefont{Lawrie}},
  \bibinfo{author}{\bibfnamefont{N.~W.} \bibnamefont{Hendrickx}},
  \bibinfo{author}{\bibfnamefont{S.~G.~J.} \bibnamefont{Philips}},
  \bibinfo{author}{\bibfnamefont{J.~S.} \bibnamefont{Clarke}},
  \bibinfo{author}{\bibfnamefont{L.~M.~K.} \bibnamefont{Vandersypen}},
  \bibnamefont{and}
  \bibinfo{author}{\bibfnamefont{M.}~\bibnamefont{Veldhorst}},
  \bibinfo{journal}{Nature} \textbf{\bibinfo{volume}{580}},
  \bibinfo{pages}{355} (\bibinfo{year}{2020}).

\bibitem[{\citenamefont{Hendrickx
  et~al.}(2020{\natexlab{a}})\citenamefont{Hendrickx, Franke, Sammak,
  Scappucci, and Veldhorst}}]{Hendrickx2020}
\bibinfo{author}{\bibfnamefont{N.~W.} \bibnamefont{Hendrickx}},
  \bibinfo{author}{\bibfnamefont{D.~P.} \bibnamefont{Franke}},
  \bibinfo{author}{\bibfnamefont{A.}~\bibnamefont{Sammak}},
  \bibinfo{author}{\bibfnamefont{G.}~\bibnamefont{Scappucci}},
  \bibnamefont{and}
  \bibinfo{author}{\bibfnamefont{M.}~\bibnamefont{Veldhorst}},
  \bibinfo{journal}{Nature} \textbf{\bibinfo{volume}{577}},
  \bibinfo{pages}{487} (\bibinfo{year}{2020}{\natexlab{a}}).

\bibitem[{\citenamefont{Hendrickx
  et~al.}(2020{\natexlab{b}})\citenamefont{Hendrickx, Lawrie, Russ, van
  Riggelen, de~Snoo, Schouten, Sammak, Scappucci, and
  Veldhorst}}]{Hendrickx2020a}
\bibinfo{author}{\bibfnamefont{N.~W.} \bibnamefont{Hendrickx}},
  \bibinfo{author}{\bibfnamefont{W.~I.~L.} \bibnamefont{Lawrie}},
  \bibinfo{author}{\bibfnamefont{M.}~\bibnamefont{Russ}},
  \bibinfo{author}{\bibfnamefont{F.}~\bibnamefont{van Riggelen}},
  \bibinfo{author}{\bibfnamefont{S.~L.} \bibnamefont{de~Snoo}},
  \bibinfo{author}{\bibfnamefont{R.~N.} \bibnamefont{Schouten}},
  \bibinfo{author}{\bibfnamefont{A.}~\bibnamefont{Sammak}},
  \bibinfo{author}{\bibfnamefont{G.}~\bibnamefont{Scappucci}},
  \bibnamefont{and}
  \bibinfo{author}{\bibfnamefont{M.}~\bibnamefont{Veldhorst}},
  \bibinfo{journal}{arXiv}  (\bibinfo{year}{2020}{\natexlab{b}}),
  \eprint{2009.04268v1}.

\bibitem[{\citenamefont{Mi et~al.}(2017)\citenamefont{Mi, Cady, Zajac, Deelman,
  and Petta}}]{Mi2017}
\bibinfo{author}{\bibfnamefont{X.}~\bibnamefont{Mi}},
  \bibinfo{author}{\bibfnamefont{J.~V.} \bibnamefont{Cady}},
  \bibinfo{author}{\bibfnamefont{D.~M.} \bibnamefont{Zajac}},
  \bibinfo{author}{\bibfnamefont{P.~W.} \bibnamefont{Deelman}},
  \bibnamefont{and} \bibinfo{author}{\bibfnamefont{J.~R.} \bibnamefont{Petta}},
  \bibinfo{journal}{Science} \textbf{\bibinfo{volume}{355}},
  \bibinfo{pages}{156} (\bibinfo{year}{2017}).

\bibitem[{\citenamefont{Samkharadze et~al.}(2018)\citenamefont{Samkharadze,
  Zheng, Kalhor, Brousse, Sammak, Mendes, Blais, Scappucci, and
  Vandersypen}}]{Samkharadze2018}
\bibinfo{author}{\bibfnamefont{N.}~\bibnamefont{Samkharadze}},
  \bibinfo{author}{\bibfnamefont{G.}~\bibnamefont{Zheng}},
  \bibinfo{author}{\bibfnamefont{N.}~\bibnamefont{Kalhor}},
  \bibinfo{author}{\bibfnamefont{D.}~\bibnamefont{Brousse}},
  \bibinfo{author}{\bibfnamefont{A.}~\bibnamefont{Sammak}},
  \bibinfo{author}{\bibfnamefont{U.~C.} \bibnamefont{Mendes}},
  \bibinfo{author}{\bibfnamefont{A.}~\bibnamefont{Blais}},
  \bibinfo{author}{\bibfnamefont{G.}~\bibnamefont{Scappucci}},
  \bibnamefont{and} \bibinfo{author}{\bibfnamefont{L.~M.~K.}
  \bibnamefont{Vandersypen}}, \bibinfo{journal}{Science}
  \textbf{\bibinfo{volume}{359}}, \bibinfo{pages}{1123} (\bibinfo{year}{2018}).

\bibitem[{\citenamefont{Li et~al.}(2018)\citenamefont{Li, Petit, Franke,
  Dehollain, Helsen, Steudtner, Thomas, Yoscovits, Singh, Wehner
  et~al.}}]{Li2018}
\bibinfo{author}{\bibfnamefont{R.}~\bibnamefont{Li}},
  \bibinfo{author}{\bibfnamefont{L.}~\bibnamefont{Petit}},
  \bibinfo{author}{\bibfnamefont{D.~P.} \bibnamefont{Franke}},
  \bibinfo{author}{\bibfnamefont{J.~P.} \bibnamefont{Dehollain}},
  \bibinfo{author}{\bibfnamefont{J.}~\bibnamefont{Helsen}},
  \bibinfo{author}{\bibfnamefont{M.}~\bibnamefont{Steudtner}},
  \bibinfo{author}{\bibfnamefont{N.~K.} \bibnamefont{Thomas}},
  \bibinfo{author}{\bibfnamefont{Z.~R.} \bibnamefont{Yoscovits}},
  \bibinfo{author}{\bibfnamefont{K.~J.} \bibnamefont{Singh}},
  \bibinfo{author}{\bibfnamefont{S.}~\bibnamefont{Wehner}},
  \bibnamefont{et~al.}, \bibinfo{journal}{Science Advances}
  \textbf{\bibinfo{volume}{4}}, \bibinfo{pages}{3960} (\bibinfo{year}{2018}).

\bibitem[{\citenamefont{Pillarisetty et~al.}(2019)\citenamefont{Pillarisetty,
  Kashani, Keys, Kotlyar, Luthi, Michalak, Millard, Roberts, Torres, Zietz
  et~al.}}]{Pillarisetty2019}
\bibinfo{author}{\bibfnamefont{R.}~\bibnamefont{Pillarisetty}},
  \bibinfo{author}{\bibfnamefont{N.}~\bibnamefont{Kashani}},
  \bibinfo{author}{\bibfnamefont{P.}~\bibnamefont{Keys}},
  \bibinfo{author}{\bibfnamefont{R.}~\bibnamefont{Kotlyar}},
  \bibinfo{author}{\bibfnamefont{F.}~\bibnamefont{Luthi}},
  \bibinfo{author}{\bibfnamefont{D.}~\bibnamefont{Michalak}},
  \bibinfo{author}{\bibfnamefont{K.}~\bibnamefont{Millard}},
  \bibinfo{author}{\bibfnamefont{J.}~\bibnamefont{Roberts}},
  \bibinfo{author}{\bibfnamefont{J.}~\bibnamefont{Torres}},
  \bibinfo{author}{\bibfnamefont{O.}~\bibnamefont{Zietz}},
  \bibnamefont{et~al.}, in \emph{\bibinfo{booktitle}{2019 {IEEE} International
  Electron Devices Meeting ({IEDM})}} (\bibinfo{publisher}{{IEEE}},
  \bibinfo{year}{2019}).

\bibitem[{\citenamefont{Wuetz et~al.}(2020)\citenamefont{Wuetz, Bavdaz, Yeoh,
  Schouten, van~der Does, Tiggelman, Sabbagh, Sammak, Almudever, Sebastiano
  et~al.}}]{Wuetz2020}
\bibinfo{author}{\bibfnamefont{B.~P.} \bibnamefont{Wuetz}},
  \bibinfo{author}{\bibfnamefont{P.~L.} \bibnamefont{Bavdaz}},
  \bibinfo{author}{\bibfnamefont{L.~A.} \bibnamefont{Yeoh}},
  \bibinfo{author}{\bibfnamefont{R.}~\bibnamefont{Schouten}},
  \bibinfo{author}{\bibfnamefont{H.}~\bibnamefont{van~der Does}},
  \bibinfo{author}{\bibfnamefont{M.}~\bibnamefont{Tiggelman}},
  \bibinfo{author}{\bibfnamefont{D.}~\bibnamefont{Sabbagh}},
  \bibinfo{author}{\bibfnamefont{A.}~\bibnamefont{Sammak}},
  \bibinfo{author}{\bibfnamefont{C.~G.} \bibnamefont{Almudever}},
  \bibinfo{author}{\bibfnamefont{F.}~\bibnamefont{Sebastiano}},
  \bibnamefont{et~al.}, \bibinfo{journal}{npj Quantum Information}
  \textbf{\bibinfo{volume}{6}} (\bibinfo{year}{2020}).

\bibitem[{\citenamefont{Pauka et~al.}(2019{\natexlab{a}})\citenamefont{Pauka,
  Das, Hornibrook, Gardner, Manfra, Cassidy, and Reilly}}]{Pauka2019}
\bibinfo{author}{\bibfnamefont{S.~J.} \bibnamefont{Pauka}},
  \bibinfo{author}{\bibfnamefont{K.}~\bibnamefont{Das}},
  \bibinfo{author}{\bibfnamefont{J.~M.} \bibnamefont{Hornibrook}},
  \bibinfo{author}{\bibfnamefont{G.~C.} \bibnamefont{Gardner}},
  \bibinfo{author}{\bibfnamefont{M.~J.} \bibnamefont{Manfra}},
  \bibinfo{author}{\bibfnamefont{M.~C.} \bibnamefont{Cassidy}},
  \bibnamefont{and} \bibinfo{author}{\bibfnamefont{D.~J.}
  \bibnamefont{Reilly}}, \bibinfo{journal}{arXiv}
  (\bibinfo{year}{2019}{\natexlab{a}}), \eprint{1908.07685v1}.

\bibitem[{\citenamefont{Borselli et~al.}(2015)\citenamefont{Borselli, Eng,
  Ross, Hazard, Holabird, Huang, Kiselev, Deelman, Warren, Milosavljevic
  et~al.}}]{Borselli2015}
\bibinfo{author}{\bibfnamefont{M.~G.} \bibnamefont{Borselli}},
  \bibinfo{author}{\bibfnamefont{K.}~\bibnamefont{Eng}},
  \bibinfo{author}{\bibfnamefont{R.~S.} \bibnamefont{Ross}},
  \bibinfo{author}{\bibfnamefont{T.~M.} \bibnamefont{Hazard}},
  \bibinfo{author}{\bibfnamefont{K.~S.} \bibnamefont{Holabird}},
  \bibinfo{author}{\bibfnamefont{B.}~\bibnamefont{Huang}},
  \bibinfo{author}{\bibfnamefont{A.~A.} \bibnamefont{Kiselev}},
  \bibinfo{author}{\bibfnamefont{P.~W.} \bibnamefont{Deelman}},
  \bibinfo{author}{\bibfnamefont{L.~D.} \bibnamefont{Warren}},
  \bibinfo{author}{\bibfnamefont{I.}~\bibnamefont{Milosavljevic}},
  \bibnamefont{et~al.}, \bibinfo{journal}{Nanotechnology}
  \textbf{\bibinfo{volume}{26}}, \bibinfo{pages}{375202}
  (\bibinfo{year}{2015}).

\bibitem[{\citenamefont{Zajac et~al.}(2016)\citenamefont{Zajac, Hazard, Mi,
  Nielsen, and Petta}}]{Zajac2016}
\bibinfo{author}{\bibfnamefont{D.}~\bibnamefont{Zajac}},
  \bibinfo{author}{\bibfnamefont{T.}~\bibnamefont{Hazard}},
  \bibinfo{author}{\bibfnamefont{X.}~\bibnamefont{Mi}},
  \bibinfo{author}{\bibfnamefont{E.}~\bibnamefont{Nielsen}}, \bibnamefont{and}
  \bibinfo{author}{\bibfnamefont{J.}~\bibnamefont{Petta}},
  \bibinfo{journal}{Physical Review Applied} \textbf{\bibinfo{volume}{6}}
  (\bibinfo{year}{2016}).

\bibitem[{\citenamefont{Veldhorst et~al.}(2017)\citenamefont{Veldhorst, Eenink,
  Yang, and Dzurak}}]{Veldhorst2017}
\bibinfo{author}{\bibfnamefont{M.}~\bibnamefont{Veldhorst}},
  \bibinfo{author}{\bibfnamefont{H.~G.~J.} \bibnamefont{Eenink}},
  \bibinfo{author}{\bibfnamefont{C.~H.} \bibnamefont{Yang}}, \bibnamefont{and}
  \bibinfo{author}{\bibfnamefont{A.~S.} \bibnamefont{Dzurak}},
  \bibinfo{journal}{Nature Communications} \textbf{\bibinfo{volume}{8}}
  (\bibinfo{year}{2017}).

\bibitem[{\citenamefont{Pauka et~al.}(2019{\natexlab{b}})\citenamefont{Pauka,
  Das, Kalra, Moini, Yang, Trainer, Bousquet, Cantaloube, Dick, Gardner
  et~al.}}]{Pauka2019a}
\bibinfo{author}{\bibfnamefont{S.~J.} \bibnamefont{Pauka}},
  \bibinfo{author}{\bibfnamefont{K.}~\bibnamefont{Das}},
  \bibinfo{author}{\bibfnamefont{R.}~\bibnamefont{Kalra}},
  \bibinfo{author}{\bibfnamefont{A.}~\bibnamefont{Moini}},
  \bibinfo{author}{\bibfnamefont{Y.}~\bibnamefont{Yang}},
  \bibinfo{author}{\bibfnamefont{M.}~\bibnamefont{Trainer}},
  \bibinfo{author}{\bibfnamefont{A.}~\bibnamefont{Bousquet}},
  \bibinfo{author}{\bibfnamefont{C.}~\bibnamefont{Cantaloube}},
  \bibinfo{author}{\bibfnamefont{N.}~\bibnamefont{Dick}},
  \bibinfo{author}{\bibfnamefont{G.~C.} \bibnamefont{Gardner}},
  \bibnamefont{et~al.}, \bibinfo{journal}{arXiv}
  (\bibinfo{year}{2019}{\natexlab{b}}), \eprint{1912.01299v1}.

\bibitem[{\citenamefont{Xu et~al.}(2020)\citenamefont{Xu, Unseld, Corna,
  Zwerver, Sammak, Brousse, Samkharadze, Amitonov, Veldhorst, Scappucci
  et~al.}}]{Xu2020}
\bibinfo{author}{\bibfnamefont{Y.}~\bibnamefont{Xu}},
  \bibinfo{author}{\bibfnamefont{F.~K.} \bibnamefont{Unseld}},
  \bibinfo{author}{\bibfnamefont{A.}~\bibnamefont{Corna}},
  \bibinfo{author}{\bibfnamefont{A.~M.~J.} \bibnamefont{Zwerver}},
  \bibinfo{author}{\bibfnamefont{A.}~\bibnamefont{Sammak}},
  \bibinfo{author}{\bibfnamefont{D.}~\bibnamefont{Brousse}},
  \bibinfo{author}{\bibfnamefont{N.}~\bibnamefont{Samkharadze}},
  \bibinfo{author}{\bibfnamefont{S.~V.} \bibnamefont{Amitonov}},
  \bibinfo{author}{\bibfnamefont{M.}~\bibnamefont{Veldhorst}},
  \bibinfo{author}{\bibfnamefont{G.}~\bibnamefont{Scappucci}},
  \bibnamefont{et~al.}, \bibinfo{journal}{arXiv}  (\bibinfo{year}{2020}),
  \eprint{2005.03851v1}.

\bibitem[{\citenamefont{Puddy et~al.}(2015)\citenamefont{Puddy, Smith, Al-Taie,
  Chong, Farrer, Griffiths, Ritchie, Kelly, Pepper, and Smith}}]{Puddy2015}
\bibinfo{author}{\bibfnamefont{R.~K.} \bibnamefont{Puddy}},
  \bibinfo{author}{\bibfnamefont{L.~W.} \bibnamefont{Smith}},
  \bibinfo{author}{\bibfnamefont{H.}~\bibnamefont{Al-Taie}},
  \bibinfo{author}{\bibfnamefont{C.~H.} \bibnamefont{Chong}},
  \bibinfo{author}{\bibfnamefont{I.}~\bibnamefont{Farrer}},
  \bibinfo{author}{\bibfnamefont{J.~P.} \bibnamefont{Griffiths}},
  \bibinfo{author}{\bibfnamefont{D.~A.} \bibnamefont{Ritchie}},
  \bibinfo{author}{\bibfnamefont{M.~J.} \bibnamefont{Kelly}},
  \bibinfo{author}{\bibfnamefont{M.}~\bibnamefont{Pepper}}, \bibnamefont{and}
  \bibinfo{author}{\bibfnamefont{C.~G.} \bibnamefont{Smith}},
  \bibinfo{journal}{Applied Physics Letters} \textbf{\bibinfo{volume}{107}},
  \bibinfo{pages}{143501} (\bibinfo{year}{2015}).

\bibitem[{\citenamefont{Schaal et~al.}(2018)\citenamefont{Schaal, Barraud,
  Morton, and Gonzalez-Zalba}}]{Schaal2018}
\bibinfo{author}{\bibfnamefont{S.}~\bibnamefont{Schaal}},
  \bibinfo{author}{\bibfnamefont{S.}~\bibnamefont{Barraud}},
  \bibinfo{author}{\bibfnamefont{J.}~\bibnamefont{Morton}}, \bibnamefont{and}
  \bibinfo{author}{\bibfnamefont{M.}~\bibnamefont{Gonzalez-Zalba}},
  \bibinfo{journal}{Physical Review Applied} \textbf{\bibinfo{volume}{9}}
  (\bibinfo{year}{2018}).

\bibitem[{\citenamefont{Schaal et~al.}(2019)\citenamefont{Schaal, Rossi,
  Ciriano-Tejel, Yang, Barraud, Morton, and Gonzalez-Zalba}}]{Schaal2019}
\bibinfo{author}{\bibfnamefont{S.}~\bibnamefont{Schaal}},
  \bibinfo{author}{\bibfnamefont{A.}~\bibnamefont{Rossi}},
  \bibinfo{author}{\bibfnamefont{V.~N.} \bibnamefont{Ciriano-Tejel}},
  \bibinfo{author}{\bibfnamefont{T.-Y.} \bibnamefont{Yang}},
  \bibinfo{author}{\bibfnamefont{S.}~\bibnamefont{Barraud}},
  \bibinfo{author}{\bibfnamefont{J.~J.~L.} \bibnamefont{Morton}},
  \bibnamefont{and} \bibinfo{author}{\bibfnamefont{M.~F.}
  \bibnamefont{Gonzalez-Zalba}}, \bibinfo{journal}{Nature Electronics}
  \textbf{\bibinfo{volume}{2}}, \bibinfo{pages}{236} (\bibinfo{year}{2019}).

\bibitem[{\citenamefont{Mills et~al.}(2019)\citenamefont{Mills, Zajac, Gullans,
  Schupp, Hazard, and Petta}}]{Mills2019}
\bibinfo{author}{\bibfnamefont{A.~R.} \bibnamefont{Mills}},
  \bibinfo{author}{\bibfnamefont{D.~M.} \bibnamefont{Zajac}},
  \bibinfo{author}{\bibfnamefont{M.~J.} \bibnamefont{Gullans}},
  \bibinfo{author}{\bibfnamefont{F.~J.} \bibnamefont{Schupp}},
  \bibinfo{author}{\bibfnamefont{T.~M.} \bibnamefont{Hazard}},
  \bibnamefont{and} \bibinfo{author}{\bibfnamefont{J.~R.} \bibnamefont{Petta}},
  \bibinfo{journal}{Nature Communications} \textbf{\bibinfo{volume}{10}}
  (\bibinfo{year}{2019}).

\bibitem[{\citenamefont{Ward et~al.}(2013)\citenamefont{Ward, Savage, Lagally,
  Coppersmith, and Eriksson}}]{Ward2013}
\bibinfo{author}{\bibfnamefont{D.~R.} \bibnamefont{Ward}},
  \bibinfo{author}{\bibfnamefont{D.~E.} \bibnamefont{Savage}},
  \bibinfo{author}{\bibfnamefont{M.~G.} \bibnamefont{Lagally}},
  \bibinfo{author}{\bibfnamefont{S.~N.} \bibnamefont{Coppersmith}},
  \bibnamefont{and} \bibinfo{author}{\bibfnamefont{M.~A.}
  \bibnamefont{Eriksson}}, \bibinfo{journal}{Applied Physics Letters}
  \textbf{\bibinfo{volume}{102}}, \bibinfo{pages}{213107}
  (\bibinfo{year}{2013}).

\end{thebibliography}
\clearpage

\pagebreak
\onecolumngrid
\section{Supplementary Information}
\renewcommand{\thefigure}{S\arabic{figure}}
\setcounter{figure}{0}
\renewcommand{\theequation}{S\arabic{equation}}
\setcounter{equation}{0}
\renewcommand{\thetable}{S\arabic{table}}
\setcounter{table}{0}

\subsection{$2^{14} \times 2^{14} \simeq 2.6 \times 10^{8}$ charge-locking units and transistor gate of 10~nm $\times$ 14~nm i.e. $C_{g} \simeq 1.4 \times 10^{-3}fF$}
For serially refreshed charge-locking array with 2D cross-bar addressing, i.e. $ N = M = 2^{14}$, the power dissipation is 
\begin{equation}
  \begin{split}
P_{S-2D} & = 5NM(N+M-\frac{8}{5})C_{g}V_{g}^2f_{c} \\
         & \simeq 5 \times 2^{28} \times 2^{15} \times 1.4 \times 10^{-18} F \times {1V}^2\times 10^{3} Hz \\
         & = 61.5mW/kHz
  \end{split}
\end{equation}

For parallel refreshed charge-locking approach with 2D addressing, i.e. $N = M = 2^{14}$ and $K = 14$, and assuming voltage variation $\delta V_{R} \simeq 100$ ~mV and $Q \simeq 10$,
\begin{equation}
  \begin{split}
P_{P-2D}& \simeq N^{2} (5 + \log_2 N + Q)C_{g}V_{g}^2f_{c} \\
        & \simeq 2^{28} \times (5 + 14+10) \times1.4 \times 10^{-18} F \times {1V}^2\times 10^{3} Hz \\
        & = 11 \mu W/kHz
\end{split}
\end{equation}

If the voltage variation $\delta V_{R}$ is improved to be around 10mV, then $Q \simeq 1$, the power dissipation can be further reduced to
\begin{equation}
  \begin{split}
P_{P-2D}& \simeq N^{2} (5 + \log_2 N + Q)C_{g}V_{g}^2f_{c} \\
        & \simeq 2^{28} \times (5 +14+1) \times1.4 \times 10^{-18} F \times {1V}^2\times 10^{3} Hz \\
        & = 7.5 \mu W/kHz
\end{split}
\end{equation}

Assume the refreshing rate is 1 kHz, the column MUX addressing gate and transistor (switch) of the charge locking unit will be switched at a rate of 100 GHz, which is not compatible with the digital electronics switching at a few GHz rate. To resolve this issue, multiple independent inputs can be used to column MUX. For example, column MUX can employ 16 independent inputs while reducing the MUX level from 14 to 10. As a result, the required switching rate is reduced from 100 GHz to 6.6 GHz met by digital electronics. The total required wire number will increase from 58 to 64. This problem is equally present in serially refreshed charge-locking array with cross-bar addressing. For the serially refreshed solution, this multiple independent inputs structure have to be employed for both column and row MUX.

\subsection{Recharging time}
Following the same process as power dissipation analysis, assuming the recharging time is limited by a simple RC time constant, where R is dominated by the output resistance of the control electronics at higher temperature.

For serially refreshed charge-locking approach with 2D cross-bar addressing,
\begin{equation}
  \begin{split}
T_{S-2D} & \propto NMC_{H} + NM max (N, M)C_{g}  + NM max (\frac{NC{g}}{2}, \frac{MC{g}}{2})  \\
  \end{split}
\end{equation}

Here the first term accounts for the time related to recharge holding capacitors. The second term accounts for the time in switching the shared control line, since the column and row shared control are simultaneously switched, the longer time is used. The third term accounts for the time in routing the DC signals i.e. switching MUX addressing gates, similarly the longer one is taken between the Column MUX and the Row MUX. For an array with equal size in each dimension, i.e. $ N= M$
\begin{equation}
  \begin{split}
T_{S-2D} & \propto N^{2}C_{H} + N^{3}C_{g} + \frac{N^{3}C_{g}}{2}  \\
        &  \propto N^{2}C_{H} + \frac{3}{2}N^{3}C_{g}
  \end{split}
\end{equation}

For parallel refreshed charge-locking array with 2D addressing, the time required to recharge a single row is 
\begin{equation}
  \begin{split}
T_{P-1D} & \propto NC_{g} + NC_{R} + KNC_{g}  \\
         & \propto NC_{g} + NC_{R} + N \log_2 N C_{g}
  \end{split}
\end{equation}

Note, there is no term related to recharging holding capacitor here, since it is not limited by RC time constant and the time required for charge redistribution to occur is assumed be at much shorter time scale.  The first term accounts for the time in switching transistors with the shared control line $GH_{0}$. The second term accounts for the time required to prepare the recharging capacitors. The third term accounts for the time in routing DC signals i.e. in switching MUX addressing gates.

Then the time required to recharge M rows in total is 
\begin{equation}
  \begin{split}
T_{P-2D} & \propto M \times T_{P-1D} + M \times \frac{MC_{g}}{2}  \\
  \end{split}
\end{equation}
The first term accounts for the time to recharge all rows of charge-locking units and the second term accounts for the time in routing DC signals by the row MUX, i.e. in switching addressing gates.
For an array with equal size in each dimension, i.e. $ N= M$
\begin{equation}
  \begin{split}
T_{P-2D} & \propto N^{2}C_{R} + N^{2}(\frac{3}{2}+ \log_2 N)C_{g}
  \end{split}
\end{equation}

Below we evaluate the recharging time, assuming realistic conditions. Suppose the transistor gate is of the size 10~nm $\times$ 14~nm, i.e. $C_{g} \simeq 10^{-6} C_{H}$. 

For any charge-locking array with size smaller than $10^{12}$, i.e. $N< 10^{6}$, the total recharging time for serially refreshed charge-locking array with 2D cross-bar addressing can be approximated as
\begin{equation}
  \begin{split}
T_{S-2D} &  \propto N^{2}C_{H} + \frac{3}{2}N^{3}C_{g} \\
         &  \propto N^{2}C_{H}
  \end{split}
\end{equation}

As the recharging capacitor $C_{R}$ is also orders of magnitude larger than the transistor capacitor $C_{H}$ in either situation, i.e. $C_{R} = 10^{-2} C_{H}$ or $C_{R} = 10^{-3} C_{H}$, the total recharging time for parallel refreshed charge-locking array can be approximated as 

\begin{equation}
  \begin{split}
T_{P-2D} & \propto N^{2}C_{R} + N^{2}(\frac{3}{2}+ \log_2 N)C_{g}\\
         & \propto N^{2}C_{R}
  \end{split}
\end{equation}

\subsection{Base-4 signal multiplexing}

\begin{figure*}[h]
\includegraphics[width=0.98 \linewidth]{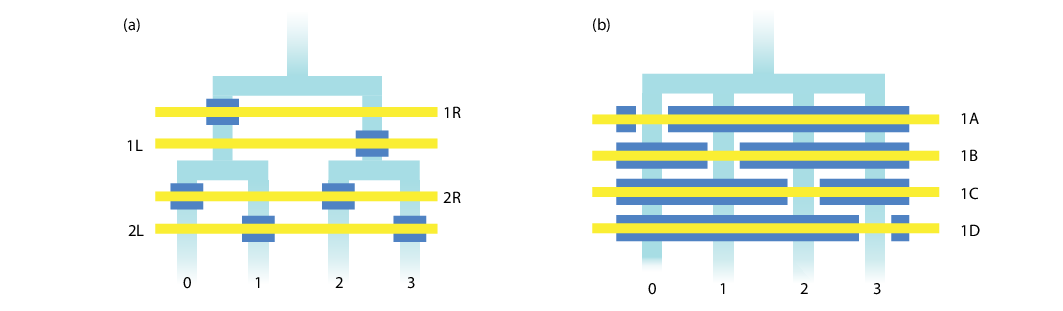}
\caption{(a) Base-2 singal multiplexing based on MSLG method, where each channel at current is split into two channels at next level. (b) Base-4 signal multipexing based on MSLG method, where each channel at current level is split into 4 channels at next level.
\label{Base-4}}
\end{figure*}

\subsection{Generalised parallel refreshed charge-locking with MLSG based MUX}
Below we show the specific sequence to operate MLSG based MUX as column MUX for parallel refreshed charge-locking, one for homogeneous base-3 and the other for heterogeneous base stacking. 

\begin{table*}
  \begin{tabular}{|p{2cm}||p{2.5cm}||p{1cm}|p{1cm}|p{1cm}|p{1cm}|p{1cm}|p{1cm}|p{1cm}|p{1cm}|p{1cm}|}
   \hline
   \hline
   Output No & Bit Map & 1A & 1B & 1C & 2A & 2B & 2C & 3A & 3B & 3C\\
   \hline
   0          & 000                &\textcolor{red}{OFF}&ON&ON &\textcolor{red}{OFF}&ON&ON &\textcolor{red}{OFF}&ON&ON\\
   9          & 100                &ON&\textcolor{red}{OFF}&ON &\textcolor{red}{OFF}&ON&ON &\textcolor{red}{OFF}&ON&ON\\
   18         & 200                &ON&ON&\textcolor{red}{OFF} &\textcolor{red}{OFF}&ON&ON &\textcolor{red}{OFF}&ON&ON\\
   21         & 210                &ON&ON&\textcolor{red}{OFF} &ON&\textcolor{red}{OFF}&ON &\textcolor{red}{OFF}&ON&ON\\
   12         & 110                &ON&\textcolor{red}{OFF}&ON &ON&\textcolor{red}{OFF}&ON &\textcolor{red}{OFF}&ON&ON\\
   3          & 010                &\textcolor{red}{OFF}&ON&ON &ON&\textcolor{red}{OFF}&ON &\textcolor{red}{OFF}&ON&ON\\
   6          & 020                &\textcolor{red}{OFF}&ON&ON &ON&ON&\textcolor{red}{OFF} &\textcolor{red}{OFF}&ON&ON\\
   15         & 120                &ON&\textcolor{red}{OFF}&ON &ON&ON&\textcolor{red}{OFF} &\textcolor{red}{OFF}&ON&ON\\
   24         & 220                &ON&ON&\textcolor{red}{OFF} &ON&ON&\textcolor{red}{OFF} &\textcolor{red}{OFF}&ON&ON\\
   25         & 221                &ON&ON&\textcolor{red}{OFF} &ON&ON&\textcolor{red}{OFF} &ON&\textcolor{red}{OFF}&ON\\
   16         & 121                &ON&\textcolor{red}{OFF}&ON &ON&ON&\textcolor{red}{OFF} &ON&\textcolor{red}{OFF}&ON\\
   7          & 021                &\textcolor{red}{OFF}&ON&ON &ON&ON&\textcolor{red}{OFF} &ON&\textcolor{red}{OFF}&ON\\
   4          & 011                &\textcolor{red}{OFF}&ON&ON &ON&\textcolor{red}{OFF}&ON &ON&\textcolor{red}{OFF}&ON\\
   13         & 111                &ON&\textcolor{red}{OFF}&ON &ON&\textcolor{red}{OFF}&ON &ON&\textcolor{red}{OFF}&ON\\
   22         & 211                &ON&ON&\textcolor{red}{OFF} &ON&\textcolor{red}{OFF}&ON &ON&\textcolor{red}{OFF}&ON\\
   19         & 201                &ON&ON&\textcolor{red}{OFF} &\textcolor{red}{OFF}&ON&ON &ON&\textcolor{red}{OFF}&ON\\
   10         & 101                &ON&\textcolor{red}{OFF}&ON &\textcolor{red}{OFF}&ON&ON &ON&\textcolor{red}{OFF}&ON\\
   1          & 001                &\textcolor{red}{OFF}&ON&ON &\textcolor{red}{OFF}&ON&ON &ON&\textcolor{red}{OFF}&ON\\
   2          & 002                &\textcolor{red}{OFF}&ON&ON &\textcolor{red}{OFF}&ON&ON &ON&ON&\textcolor{red}{OFF}\\
   11         & 102                &ON&\textcolor{red}{OFF}&ON &\textcolor{red}{OFF}&ON&ON &ON&ON&\textcolor{red}{OFF}\\
   20         & 202                &ON&ON&\textcolor{red}{OFF} &\textcolor{red}{OFF}&ON&ON &ON&ON&\textcolor{red}{OFF}\\
   23         & 212                &ON&ON&\textcolor{red}{OFF} &ON&\textcolor{red}{OFF}&ON &ON&ON&\textcolor{red}{OFF}\\
   14         & 112                &ON&\textcolor{red}{OFF}&ON &ON&\textcolor{red}{OFF}&ON &ON&ON&\textcolor{red}{OFF}\\
   5          & 012                &\textcolor{red}{OFF}&ON&ON &ON&\textcolor{red}{OFF}&ON &ON&ON&\textcolor{red}{OFF}\\
   8          & 022                &\textcolor{red}{OFF}&ON&ON &ON&ON&\textcolor{red}{OFF} &ON&ON&\textcolor{red}{OFF}\\
   17         & 122                &ON&\textcolor{red}{OFF}&ON &ON&ON&\textcolor{red}{OFF} &ON&ON&\textcolor{red}{OFF}\\
   26         & 222                &ON&ON&\textcolor{red}{OFF} &ON&ON&\textcolor{red}{OFF} &ON&ON&\textcolor{red}{OFF}\\
   \hline
 \end{tabular}
 \caption{Sequence to operate a 3-level base-3 MUX for parallel refreshed charge-locking. Inactive addressing gates are labelled as OFF and active addressing gates are labelled as ON. For MUX of homogeneous base, there is a simple correspondence between the output channel and the inactive addressing gates. For example, 23 corresponds to 212 that is $23 = 2 \times 3^{2} + 1 \times 3^{1} + 2 \times 2^{0}$ . Physically means all addressing gates other than gate C of level 1, gate B of level 2 and gate C of level 3 are activated to select output channel 23. Upon the transition from 18 to 21, 2nd level inactive addressing gate is changed from A to B. The activated addressing gate A of 2nd level keeps output channel 0, 9, 18 or equivalently 000, 100, 200 in charge-locked state. Upon the transition from 24 to 25, 3rd level inactive addressing is changed from A to B. The activated addressing gate A of 3rd level keeps output channel 0, 9, 18, 21, 12, 3, 6, 15, 24 or equivalently 000, 100, 200 210, 110, 010, 020, 120, 220 in charge-locked state.}
\end{table*}

\begin{table*}
  \begin{tabular}{|p{2cm}||p{2cm}||p{1cm}|p{1cm}|p{1cm}|p{1cm}|p{1cm}|p{1cm}|p{1cm}|p{1cm}|p{1cm}|p{1cm}|}
   \hline
   \hline
   Output No & Bit Map & 1A & 1B & 2A & 2B & 2C & 2D & 2E & 3A & 3B & 3C\\
   \hline
   0          & 000                &\textcolor{red}{OFF}&ON  &\textcolor{red}{OFF}&ON&ON&ON&ON   &\textcolor{red}{OFF}&ON&ON\\
   15         & 100                &ON&\textcolor{red}{OFF}  &\textcolor{red}{OFF}&ON&ON&ON&ON   &\textcolor{red}{OFF}&ON&ON\\
   18         & 110                &ON&\textcolor{red}{OFF}  &ON&\textcolor{red}{OFF}&ON&ON&ON   &\textcolor{red}{OFF}&ON&ON\\
   3          & 010                &\textcolor{red}{OFF}&ON  &ON&\textcolor{red}{OFF}&ON&ON&ON   &\textcolor{red}{OFF}&ON&ON\\
   6          & 020                &\textcolor{red}{OFF}&ON  &ON&ON&\textcolor{red}{OFF}&ON&ON   &\textcolor{red}{OFF}&ON&ON\\
   21         & 120                &ON&\textcolor{red}{OFF}  &ON&ON&\textcolor{red}{OFF}&ON&ON   &\textcolor{red}{OFF}&ON&ON\\
   24         & 130                &ON&\textcolor{red}{OFF}  &ON&ON&ON&\textcolor{red}{OFF}&ON   &\textcolor{red}{OFF}&ON&ON\\
   9          & 030                &\textcolor{red}{OFF}&ON  &ON&ON&ON&\textcolor{red}{OFF}&ON   &\textcolor{red}{OFF}&ON&ON\\
   12         & 040                &\textcolor{red}{OFF}&ON  &ON&ON&ON&ON&\textcolor{red}{OFF}   &\textcolor{red}{OFF}&ON&ON\\
   27         & 140                &ON&\textcolor{red}{OFF}  &ON&ON&ON&ON&\textcolor{red}{OFF}   &\textcolor{red}{OFF}&ON&ON\\
   28         & 141                &ON&\textcolor{red}{OFF}  &ON&ON&ON&ON&\textcolor{red}{OFF}   &ON&\textcolor{red}{OFF}&ON\\
   13         & 041                &\textcolor{red}{OFF}&ON  &ON&ON&ON&ON&\textcolor{red}{OFF}   &ON&\textcolor{red}{OFF}&ON\\
   10         & 031                &\textcolor{red}{OFF}&ON  &ON&ON&ON&\textcolor{red}{OFF}&ON   &ON&\textcolor{red}{OFF}&ON\\
   25         & 131                &ON&\textcolor{red}{OFF}  &ON&ON&ON&\textcolor{red}{OFF}&ON   &ON&\textcolor{red}{OFF}&ON\\
   22         & 121                &ON&\textcolor{red}{OFF}  &ON&ON&\textcolor{red}{OFF}&ON&ON   &ON&\textcolor{red}{OFF}&ON\\
   7          & 021                &\textcolor{red}{OFF}&ON  &ON&ON&\textcolor{red}{OFF}&ON&ON   &ON&\textcolor{red}{OFF}&ON\\
   4          & 011                &\textcolor{red}{OFF}&ON  &ON&\textcolor{red}{OFF}&ON&ON&ON   &ON&\textcolor{red}{OFF}&ON\\
   19         & 111                &ON&\textcolor{red}{OFF}  &ON&\textcolor{red}{OFF}&ON&ON&ON   &ON&\textcolor{red}{OFF}&ON\\
   16         & 101                &ON&\textcolor{red}{OFF}  &\textcolor{red}{OFF}&ON&ON&ON&ON   &ON&\textcolor{red}{OFF}&ON\\
   1          & 001                &\textcolor{red}{OFF}&ON  &\textcolor{red}{OFF}&ON&ON&ON&ON   &ON&\textcolor{red}{OFF}&ON\\
   2          & 002                &\textcolor{red}{OFF}&ON  &\textcolor{red}{OFF}&ON&ON&ON&ON   &ON&ON&\textcolor{red}{OFF}\\
   17         & 102                &ON&\textcolor{red}{OFF}  &\textcolor{red}{OFF}&ON&ON&ON&ON   &ON&ON&\textcolor{red}{OFF}\\
   20         & 112                &ON&\textcolor{red}{OFF}  &ON&\textcolor{red}{OFF}&ON&ON&ON   &ON&ON&\textcolor{red}{OFF}\\
   5          & 012                &\textcolor{red}{OFF}&ON  &ON&\textcolor{red}{OFF}&ON&ON&ON   &ON&ON&\textcolor{red}{OFF}\\
   8          & 022                &\textcolor{red}{OFF}&ON  &ON&ON&\textcolor{red}{OFF}&ON&ON   &ON&ON&\textcolor{red}{OFF}\\
   23         & 122                &ON&\textcolor{red}{OFF}  &ON&ON&\textcolor{red}{OFF}&ON&ON   &ON&ON&\textcolor{red}{OFF}\\
   26         & 132                &ON&\textcolor{red}{OFF}  &ON&ON&ON&\textcolor{red}{OFF}&ON   &ON&ON&\textcolor{red}{OFF}\\
   11         & 032                &\textcolor{red}{OFF}&ON  &ON&ON&ON&\textcolor{red}{OFF}&ON   &ON&ON&\textcolor{red}{OFF}\\
   14         & 042                &\textcolor{red}{OFF}&ON  &ON&ON&ON&ON&\textcolor{red}{OFF}   &ON&ON&\textcolor{red}{OFF}\\
   29         & 142                &ON &\textcolor{red}{OFF}  &ON&ON&ON&ON&\textcolor{red}{OFF}   &ON&ON&\textcolor{red}{OFF}\\
   \hline
 \end{tabular}
 \caption{Sequence to operate a 3-level MUX of heterogeneous base (1st level of base2, 2nd level of base 5 and 3rd level of base 3) for parallel refreshed charge-locking. Inactive addressing gates are labelled as OFF and active addressing gates are labelled as ON. There is a one to one correspondence between the inactive addressing gates and the output channel. For example, output channel 26 is selected for charging by activating all addressing gates other than gate B of 1st level, gate D of 2nd level and gate C of 3rd level, that is $26 = 1 \times (3 \times 5) + 3 \times 3 + 1 \times 2$.  Upon the transition from 15 to 18, 2nd level inactive addressing gate is changed from A to B. The activated addressing gate A of 2nd level keeps output channel 0, 15 or equivalently 000, 100 in charge-locked state. Upon the transition from 27 to 28, 3rd level inactive addressing is changed from A to B. The activated addressing gate 0 of 3rd level keeps output channel 0, 15, 18, 3, 6, 21, 24, 9, 12 or equivalently 000, 100, 110 010, 020, 120, 130, 030, 040, 140 in charge-locked state. }
\end{table*}

\end{document}